\documentclass[11pt]{article}

\usepackage{amssymb}
\usepackage{soul}
\usepackage{textcomp}
\usepackage{gensymb}
\usepackage[geometry]{ifsym}

\usepackage{latexsym}
\usepackage{bm}
\usepackage[fleqn]{amsmath}
\usepackage{graphics}
\usepackage{subfigure}
\usepackage{epsfig}
\usepackage[table,usenames,dvipsnames]{xcolor}
\usepackage{overpic}
\usepackage{tikz}
\usepackage{etoolbox}

\usepackage{url}
\usepackage{psfrag}
\usepackage{graphicx}
\usepackage{pstool}
\usepackage[title]{appendix}
\usepackage{float}
\usepackage{placeins}

\newcommand{\lowersym}[1]{\protect\raisebox{-0.45ex}{#1}}

\usepackage[noblocks]{authblk}
\usepackage[margin=1in]{geometry}
\usepackage[title]{appendix}
\usepackage[numbers]{natbib}

\usepackage{caption}
  \DeclareTextFontCommand\textsfi{\usefont{OT1}{cmss}{m}{sl}}
  \DeclareMathAlphabet\mathsfi            {OT1}{cmss}{m}{sl}
  \DeclareTextFontCommand\textsfb{\usefont{OT1}{cmss}{bx}{n}}
  \DeclareMathAlphabet\mathsfb            {OT1}{cmss}{bx}{n}
  \DeclareTextFontCommand\textsfbi{\usefont{OT1}{cmss}{m}{sl}}
  \DeclareMathAlphabet\mathsfbi            {OT1}{cmss}{m}{sl}
  \IfFileExists{t1phv.fd}
  {\DeclareTextFontCommand\textsfbi{\usefont{T1}{phv}{b}{it}}
  \DeclareMathAlphabet\mathsfbi            {T1}{phv}{b}{it}
  }{}
  \IfFileExists{ot1phv.fd}
  {\DeclareTextFontCommand\textsfbi{\usefont{OT1}{phv}{b}{it}}
  \DeclareMathAlphabet\mathsfbi            {OT1}{phv}{b}{it}
  }{}

\newcommand\affiliation[1]{\gdef\@affiliation{\let\aff\aff@inst#1}}
\gdef\@affiliation{}

\def\aff#1{\ignorespaces\textsuperscript{#1}}

\numberwithin{equation}{section}

\renewenvironment{abstract}
{\begin{quote}
\noindent \rule{\linewidth}{.5pt}\par{\bfseries \abstractname.}}
{\medskip\noindent \rule{\linewidth}{.5pt}
\end{quote}
}

\usepackage{hyperref}

\definecolor{darkblue}{rgb}{0,0,0.80}

\hypersetup{
    colorlinks=true,
    linkcolor={darkblue},
    citecolor={darkblue},
    filecolor=black,
    urlcolor={darkblue},
    plainpages=false,
}

\begin{document}






\title{\bf Recursive one-way Navier Stokes equations \\ with PSE-like cost}





\author{\bf Min Zhu}
\author{\bf Aaron Towne}

\affil{\normalsize Department of Mechanical Engineering, University of Michigan, Ann Arbor, MI, USA}

\date{}
\setcounter{Maxaffil}{1}


 \maketitle

\vspace{-1cm}


\begin{abstract}
Spatial marching methods, in which the flow state is spatially evolved in the downstream direction, can be used to produce low-cost models of flows containing a slowly varying direction, such as mixing layers, jets, and boundary layers.  The parabolized stability equations (PSE) are popular due to their extremely low cost but can only capture a single instability mode; all other modes are damped or distorted by regularization methods required to stabilize the spatial march, precluding PSE from properly capturing non-modal behavior, acoustics, and interactions between multiple instability mechanisms.  The one-way Navier-Stokes (OWNS) equations properly retain all downstream-traveling modes at a cost that is a fraction of that of global methods but still one to two orders of magnitude higher than PSE.  In this paper, we introduce a new variant of OWNS whose cost, both in terms of CPU time and memory requirements, approaches that of PSE while still properly capturing the contributions of all downstream-traveling modes.  The method is formulated in terms of a projection operator that eliminates upstream-traveling modes.  Unlike previous OWNS variants, the action of this operator on a vector can be efficiently approximated using a series of equations that can be solved recursively, i.e., successively one after the next, rather than as a coupled set.  In addition to highlighting the improved cost scaling of our method, we derive explicit error expressions and elucidate the relationship with previous OWNS variants.  The properties, efficiency, and accuracy of our method are demonstrated for both free-shear and wall-bounded flows.  
\end{abstract}






 \section{Introduction}
 \label{Sec:intro}

Despite the inherent nonlinearity of the Navier-Stokes equations, models based on linearized flow equations have proven useful for understanding physical mechanisms and predicting key flow features and behaviors.  For example, linear stability and transient growth analyses have provided insight into receptivity of laminar flows to small disturbances and the ensuing transition to turbulence \cite{Schmid2000stability}.  Linear models have also been used to study energy amplification mechanisms within turbulent flows \cite{Jovanovic2005componentwise, McKeon2010critical} and model the resulting coherent flow structures \cite{Towne2018spectral, Schmidt2018spectral}.  

A straightforward implementation of linear analyses involves discretizing the linearized Navier-Stokes (LNS) equations in each inhomogeneous spatial dimension, often referred to as a global analysis.  Global linear analysis has been fruitfully used to study a wide range of flows, including boundary layers \cite{Nichols2017stability}, cavity flow \cite{Meseguer2014linear, Citro2015linear}, wakes \cite{Barkley2006linear}, and jets \cite{Nichols2011global, Schmidt2018spectral, Edgington2021waves}, among others.  Nevertheless, the CPU and memory costs of these computations remains substantial even for flows with just two inhomogeneous directions, while the cost of fully three dimensional linear analyses approach that of the corresponding nonlinear simulations. Accordingly, there is substantial interest in reducing the cost of linear modeling methods.

The slow variation of many flows of interest in the streamwise direction, e.g., the direction tangent to a developing boundary layer or jet, offers an opportunity to reduce the cost of model development.  For example, the evolution of small disturbance in a boundary layer is central to its laminar-turbulence transition and the dynamics of coherent wavepacket structures are responsible for peak noise emission from turbulent jets.  The classical approach to leveraging this slow variation is to independently consider the stability of each location along the streamwise direction as part of a locally parallel analysis \cite{Crighton1976stability} or to stitch these local solutions together \emph{ex post facto} within a weakly nonparallel analysis \cite{Huerre1990local}.  While these approaches can provide critical insight into the types and behaviors of linear modes within a particular flow \cite{Towne2017acoustic}, they are limited in their ability to capture the evolution of a mode as it propagates or the interaction of multiple modes.  

The parabolized stability equations seek to overcome these limitations by tracking the downstream evolution of a disturbance specified at some initial location.  Using an ansatz similar to earlier weakly nonparallel methods, PSE divides the flow state into a rapidly varying wave-like component and a slowly varying shape-function at each frequency.  Paired with a constraint meant to push as much of the streamwise variation of the solution into the wave-like component, the linearized equations governing the shape-function are then integrated in the streamwise direction.  PSE is typically several orders of magnitude faster than global methods, and both linear and non-linear variants of PSE have been successfully used to study a variety of slowly varying wall-bounded \cite{bertolotti1991analysis,malik1994crossflow} and free shear \cite{day2001nonlinear,gudmundsson2011instability} flows.

Despite its success in some scenarios, PSE suffers from several limitations, which stem from the fact that, despite their name, the equations governing the shape-function are not parabolic.  This is a consequence of the inherent boundary-value nature of Navier-Stokes equations \cite{Kreiss2004initial}, and consequently the LNS operator supports modes that transfer energy in both the upstream and downstream directions, which we call leftgoing and rightgoing modes, respectively.  For well-posedness, the values of rightgoing and leftgoing modes should be specified at the upstream and downstream boundaries of the domain, respectively.  It is thus ill-posed to integrate the LNS or PSE equations in the downstream direction, and doing so causes decaying leftgoing waves to be wrongly interpreted as growing rightgoing waves, leading to instability of the march \cite{towne2019critical}.

Stability of the PSE march is achieved by applying one of several available regularization methods.  The most common technique relies on numerical dissipation introduced by using an implicit Euler integration scheme along with a \emph{minimum} step-size restriction to ensure sufficient dissipation.  Alternative methods include the explicit addition of a damping term to to the PSE evolution equation \cite{Andersson1998on} or neglecting the streamwise pressure gradient \cite{Chang1991compressible, Li1996on}.  These methods successfully stabilize the spatial march but create unintended consequences -- in addition to damping the unwanted leftgoing modes, they damp and/or distort all but one of the rightgoing waves \cite{towne2019critical}.  The single wave that is well-captured by PSE usually corresponds to the most unstable mode of the flow, e.g.,  the Kelvin-Helmholtz instability for free-shear flows and Tollmien–Schlichting or Mack modes for subsonic and supersonic wall-bounded flows, respectively.  The mode to be tracked is selected by specifying its locally parallel eigenfunction and eigenvalue as initial conditions at the beginning of the spatial march.  Consequently, PSE can give an accurate solution for flows dominated by a single instability mode, but significant errors are created for applications where non-modal or multiple modal instabilities are important \cite{cheung_lele_2009,Towne2013improved}.  


The one-way Navier-Stokes (OWNS) equations offer a well-posed alternative spatial marching method capable of capturing all rightgoing modes \cite{Towne2015one}.  This is accomplished by deriving a `one-way' operator that governs the spatial evolution of the same rightgoing modes supported by the LNS operator but does not support any leftgoing modes.  As a result, integrating the one-way equations in the streamwise direction is well-posed, enabling a stable march without need for detrimental regularization, allowing the contribution of all rightgoing modes to be properly captured.  This one-way operator is derived in terms of the eigenvectors and eigenvalues of the LNS operator, but the cost of computing this eigen-decomposition, potentially at every step in the spatial march, is intolerably high and on par with that of solving the full global problem.  This cost is drastically reduced by using ideas originally developed for constructing high-order non-reflecting outflow boundary conditions to replace the exact one-way operator with an approximation defined in terms of a set of recursion equations.  The approximate one-way equation is also formally well-posed, converges to the exact operator as the order of the recursion equations increases, and is typically an order of magnitude faster than global methods while achieving a similar level of accuracy.  This form of OWNS has come to be known as OWNS-O due to its connection with outflow boundary conditions.  

A limitation of OWNS-O is that it cannot accommodate a forcing term applied to the linearized equations, which can be used to model the impact of nonlinear terms or external inputs such as control actuation.  This lead to the development of a second variant of OWNS that is formulated in terms a projection operator that, when applied to the flow state vector, removes leftgoing modes without altering rightgoing modes \cite{towne2016advancements}.  This projection operator can be applied to the LNS equations to remove support for leftgoing modes and obtain a well-posed one-way evolution equation for the rightgoing modes.  The projection is also applied to any desired forcing term to ensure that it excites only rightgoing modes.  This capability has, for example, been used to efficiently approximate singular modes of the resolvent operator \cite{Towne2021fast, Rigas2021fast}, which provide a useful approximation of coherent flow structures \cite{McKeon2010critical, Towne2018spectral}, in slowly varying flows.  As was the case for OWNS-O, this projection-based form of OWNS is derived in terms of eigenvalues and eigenvectors of the LNS operator, but efficiency is achieved using a set of recursion equations whose solution approximates the action of the exact projection operator on a vector.  Since it is formulated in terms of a projection operator, this form of OWNS is known as OWNS-P.


OWNS-O and OWNS-P are typically an order of magnitude less costly than global linear methods in terms of both speed and memory requirements.  At the same time, OWNS is still one to two orders of magnitude more costly than PSE in both categories.  Moreover, the additional capabilities of OWNS-P come at the expense of efficiency -- OWNS-P is around three times slower than OWNS-O and requires at least twice the memory.  The dominant expense in both variants of OWNS is solving the recursion equations that remove leftgoing modes at each step in the march.  While we call these recursion equations following the terminology of the non-reflecting-boundary-conditions community \cite{Hagstrom2004new}, they cannot, in fact, be solved recursively, i.e., they cannot be solved successively one after the next.  Instead, the full set of equation is coupled due to asymmetries in their terminal conditions and must be solved all together as a much larger system of equations.   This unfortunate property limits the cost benefits of OWNS-O and OWNS-P compared to LNS and yields the higher cost relative to PSE.

In this paper, we develop a new variant of OWNS whose cost approaches that of PSE.  Like OWNS-P, our new method is formulated in terms of a projection operator and therefore is suitable for applications that necessitate a forcing on the linearized equations, e.g., approximating resolvent modes, computing the response to an exogenous input, or retaining nonlinear terms as in nonlinear PSE.  The key difference compared to OWNS-P (or OWNS-O) is that the projection operator is approximated using recursion equations that \emph{can} be solved recursively, i.e., successively one after the next, rather than as a coupled set.  This significantly reduces both the CPU and memory cost of the method.  Due to the recursive nature of our new OWNS method, we give it the name OWNS-R.  

Given their common reliance on a projection operator that splits the state into rightgoing and leftgoing modes, we explore the similarities and differences between OWNS-P and OWNS-R.  We show that the two methods produce different convergent approximations of the same projection operator.  Whereas the OWNS-P projection operator shares the same eigenvalues as the exact projection operator and approximates its eigenvectors, the OWNS-R projection operator approximates the eigenvalues but has the same eigenvectors as the exact projection operator.  A useful advantage of exact eigenvectors is that it enables derivation of an exact expression for the OWNS-R projection error for each LNS mode as a function only of its eigenvalue and the choice of recursion parameters, allowing  \emph{a priori} error assessment using a rough guess of the LNS eigenvalues and no knowledge of the eigenvectors.  An analogous result does not exist for OWNS-O or OWNS-P.  Critically, the convergence of the OWNS-R approximation depends on the same criteria as OWNS-O and OWNS-P, so recursion parameters that have been developed for a variety of flows for these previous OWNS variants can also be used for OWNS-R.

We also suggest an alternative approach for deploying the projection operator. Rather than applying it to the LNS evolution equation to obtain a one-way equation that is subsequently integrated as in OWNS-P, we simply apply the projection operator to the state vector to remove the contribution of leftgoing modes after each step in the march as the unmodified LNS operator is integrated.  We show that this simplified implementation is equivalent or nearly equivalent to the previous approach (depending on integration scheme), which ensures that the method remains well-posed and stable on a discrete level.  An advantage of our simplified approach is that it eliminates the need to work in terms of characteristic variables, making it easier to incorporate into existing codes and lowering the barrier of entry for potential users.  

The remainder of the paper is organized as follows.  In Section~\ref{sec:method}, we formulate the OWNS-R method, discuss its implementation and cost scaling, and derive error expressions.  Detailed comparisons between OWNS-P and OWNS-R are made in Section~\ref{Comparison to OWNS-P}.  Section~\ref{sec:examples} contains three example applications: a simple acoustics problem used to illustrate properties of the method, a turbulent jet, and a supersonic boundary layer.  Finally, we summarize the paper and discuss future developments in Section~\ref{sec:conclusions}.

 
\section{Method}
\label{sec:method}


\subsection{Problem formulation}
We begin with the compressible Navier-Stokes equations (including mass, momemntum, and energy equations), written schematically as
\begin{equation}\label{Equation: Abstract Navier Stokes equations}
\frac{\partial q}{\partial t} = \mathcal{N}(q).
\end{equation}
The state vector $q(x,y,z,t)\in R^{N_q}$ contains velocity components and two thermodynamics variales, e.g., density and pressure, and  $N_q$ is the number of the state variables, i.e., $N_q = 5$ for three-dimensional flows.

The Navier-Stokes operator $\mathcal{N}$ in (\ref{Equation: Abstract Navier Stokes equations}) is nonlinear.  Applying the Reynolds decomposition,
\begin{equation}\label{reynolds decomposition}
    q(x,y,z,t) = \overline{q}(x,y,z)+q^{\prime}(x,y,z,t),
\end{equation}
and moving terms that are linear and nonlinear in the perturbation $q^{\prime}$ to the left-hand-side and right-hand-side, respectively, results in an equation of the form
\begin{equation}\label{Equation: linearized Navier Stokes equations}
    \frac{\partial q^{\prime}}{\partial t} + L q^{\prime} = f,
\end{equation}
where
\begin{equation}
    \label{Eq:L_def}
    {L}(\overline{q}) = \frac{\partial \mathcal{N}}{\partial q} \bigg\rvert_{\bar{q}}
\end{equation}
is the linearized Navier-Stokes operator and $f=f(\overline{q},q^{\prime})$ contains the remaining nonlinear terms, which we view as a forcing on the linearized equations. More generally, $f$ may contain both the nonlinear terms as well as exogenous forcing applied to the flow stemming from boundary conditions, control actuation, and so on. 

To enable development of a one-way equation, we separate the linear operator into two parts, 
\begin{equation}
    \label{Eq:L_eq_Adx_B}
    L = A\frac{\partial}{\partial x}+B.    
\end{equation}
Here, $x$ is the coordinate direction in which the mean flow is slowly varying, and thus the direction in which we wish to spatially integrate the equations.  In~(\ref{Eq:L_eq_Adx_B}), we have isolated the derivatives in the $x$ direction, and all other terms from $L$ are in the operator $B$. The linear operator ${L}$ also contains the second derivative in $x$ arising from the viscous terms in the Navier-Stokes equations.  While not strictly necessary \cite{towne2016advancements,kamal2020application}, we neglect these terms in the present formulation for the sake of simplicity and note that extensive testing has shown them to be unimportant in a variety of flows \cite{Towne2015one}. Substituting~(\ref{Eq:L_eq_Adx_B}) into~(\ref{Equation: linearized Navier Stokes equations}), we have
\begin{equation}\label{Equation: linearized Navier Stokes equations 2}
    \frac{\partial q^{\prime}}{\partial t}+{A}\frac{\partial q^{\prime}}{\partial x}+{B}q^{\prime} = f.
\end{equation}

Discretizing~(\ref{Equation: linearized Navier Stokes equations 2}) in the transverse directions, i.e., $y$ and $z$, using a collocation method such as finite differences leads to the the semi-discrete form
\begin{equation}\label{Equation: discretized linearized Navier Stokes equations}
    \frac{\partial \boldsymbol{q}^{\prime}}{\partial t}+\boldsymbol{A}\frac{\partial \boldsymbol{q}^{\prime}}{\partial x}+\boldsymbol{B}\boldsymbol{q}^{\prime}=\boldsymbol{f}.
\end{equation}
Here, $\boldsymbol{q}^{\prime}\in R^{N}$ and $\boldsymbol{A}, \boldsymbol{B}\in R^{N \times N}$ are the semi-discrete forms of $q^{\prime}$, $A$, and $B$, respectively, and $N=N_q \times N_d$ is the size of the discretized state, where $N_d$ is the total number of discretization points in the transverse directions. 

Previous OWNS variants proceed by transforming the state vector $\boldsymbol{q}^{\prime}$ and its governing equation~(\ref{Equation: linearized Navier Stokes equations 2}) into characteristic space, which aids in developing recursion equations that lead to efficient implementations of the one-way equations.  While this is not necessary for the current formulation, we follow this precedent to enable straightforward comparisons with these earlier variants.  Characteristic variables are obtained via the transformation
\begin{equation}
    \boldsymbol{\phi}(x,t)=\boldsymbol{T}(x)\boldsymbol{q}^{\prime}(x,t),
\end{equation}
where $\boldsymbol{T}$ is a matrix containing the eigenvectors of $\boldsymbol{A}$,
\begin{equation}
    \boldsymbol{TAT}^{-1}=\tilde{\boldsymbol{A}},
\end{equation}
and $\tilde{\boldsymbol{A}}$ is a diagonal matrix containing the eigenvalues. Then (\ref{Equation: discretized linearized Navier Stokes equations}) can be written in terms of the characteristic variables as
\begin{equation}\label{characteristic hyperbolic equation }
    \frac{\partial\boldsymbol{\phi}}{\partial t}+\tilde{\boldsymbol{A}}\frac{\partial\boldsymbol{\phi}}{\partial x}+\tilde{\boldsymbol{B}}\boldsymbol{\phi}=\boldsymbol{f}_{\phi},
\end{equation}
where $\boldsymbol{f}_{\phi}=\boldsymbol{T}\boldsymbol{f}$ and $\tilde{\boldsymbol{B}}=\boldsymbol{TBT}^{-1}+\tilde{\boldsymbol{A}}\boldsymbol{T}\frac{\mathrm{d}\boldsymbol{T}^{-1}}{\mathrm{d}x}$.

Our goal is to obtain a one-way equation in the frequency (Fourier) domain.  However, we begin by applying a more general Laplace transform in time, which enables a rigorous identification of rightgoing and leftgoing waves.  Taking a Laplace transform of~(\ref{characteristic hyperbolic equation }) gives
\begin{equation}\label{laplace transform of characteristic hyperbolic equation}
    s\hat{\boldsymbol{\phi}}+\tilde{\boldsymbol{A}}\frac{\partial\hat{\boldsymbol{\phi}}}{\partial x}+\tilde{\boldsymbol{B}}\hat{\boldsymbol{\phi}}=\hat{\boldsymbol{f}}_{\phi},
\end{equation}
where $\hat{\boldsymbol{\phi}}(x,s)$ is the Laplace transform of $\boldsymbol{\phi}(x,t)$, $s = \eta -i\omega$, $i$ is the imaginary unit, and $\eta$ and $\omega$ are real scalars. We will later set $\eta = 0$ to revert to the Fourier domain.  Solving for $x$ derivatives, (\ref{laplace transform of characteristic hyperbolic equation}) can be written as
\begin{equation}\label{Equation: spatial marching equation}
    \frac{\mathrm{d}\hat{\boldsymbol{\phi}}}{\mathrm{d}x}=\boldsymbol{M}(x,s)\hat{\boldsymbol{\phi}}+\hat{\boldsymbol{g}}(x,s)
\end{equation}
with
\begin{subequations}
\begin{align}
     &\boldsymbol{M}=-\tilde{\boldsymbol{A}}^{-1}(s\boldsymbol{I}+\tilde{\boldsymbol{B}}),\label{definition of M} \\
    &\hat{\boldsymbol{g}}=\tilde{\boldsymbol{A}}^{-1}\hat{\boldsymbol{f}}_{\phi}. \label{definition of h}
\end{align}
\end{subequations}
Here, we have assumed that $\boldsymbol{A}$, and thus $\tilde{\boldsymbol{A}}$, is full rank.  A simple procedure for accommodating a singular $\boldsymbol{A}$ is presented in Appendix~\ref{App:singularA}, and we note that handling this case is considerably more straightforward within the OWNS-R formulation than in OWNS-P \cite{Towne2021fast}.  

The manipulations leading from (\ref{Equation: Abstract Navier Stokes equations}) to (\ref{Equation: spatial marching equation}) have produced a form more conducive to the analysis to follow but have not altered the basic character of the equations.  Indeed, a global solution of the linearized equations could be obtained by discretizing (\ref{Equation: spatial marching equation}) in the $x$ direction, applying boundary conditions at the begining and end of the $x$ domain, and solving the resulting linear system of equations.  However, this approach is computationally costly and does not leverage the slow variation of the mean flow.  Instead, the solution of (\ref{Equation: spatial marching equation}) can be obtained by spatial integration. While integrating (\ref{Equation: spatial marching equation}) is nominally ill-posed and numerically unstable, these issues can be overcome by removing leftgoing components of the solution that are responsible for the ill-posedness and instability. 

To this end, consider the eigen-decomposition of the spatial marching operator,
\begin{equation}\label{Equation: eigendecomposition of M}
    \boldsymbol{M}=\boldsymbol{VDV}^{-1} = \boldsymbol{V}i\boldsymbol{\Lambda} \boldsymbol{V}^{-1}.
\end{equation}
The eigenvalues and eigenvectors of $\boldsymbol{M}$ are located in the columns of $\boldsymbol{V}$ and in the diagonal matrix $\boldsymbol{\Lambda}$, respectively.  Here, we have assumed that $\boldsymbol{M}$ has a full basis of eigenvectors; the defective case is discussed in detail by Towne et al. \cite{Towne2015one}. The solution of (\ref{Equation: spatial marching equation}) can be written as an expansion in the eigenvectors of $\boldsymbol{M}$,
\begin{equation}\label{Equation: linear combination1}
    \hat{\boldsymbol{\phi}}(x) = \sum_{i=1}^{N}{\boldsymbol{v}_k(x)\psi_k(x)} =\boldsymbol{V\psi},
\end{equation}
where $\boldsymbol{v}_k$ is the $k$-th eigenvector of $\boldsymbol{M}$, $\psi_{k}$ is the associated expansion coefficient that determines the contribution of the eigenvector to the state vector, and $\boldsymbol{\psi} \in C^N$ is a vector containing all of the expansion coefficients.

The solution contains contributions from both rightgoing modes and leftgoing modes. Briggs criteria \cite{Briggs1964electron}, and its extension to non-parallel flows \cite{Kreiss1970initial, Huerre1990local, Towne2015one}, provides a rigorous means to distinguish between rightgoing and leftgoing modes. 
Denote $\alpha_{k}(x,s)$ as the eigenvalue corresponding to the eigenvector $\boldsymbol{v}_k$, i.e., the $k$-th element in $\boldsymbol{\Lambda}$. Then the $k$-th mode is rightgoing at $x = x_0$ if
\begin{equation}\label{Briggs, rightgoing}
    \lim_{\eta\to +\infty}\text{Im}[\alpha_k(x_0,s)]= +\infty 
\end{equation}
and leftgoing at $x = x_0$ if 
\begin{equation}\label{Briggs, leftgoing}
    \lim_{\eta\to +\infty}\text{Im}[\alpha_k(x_0,s)]= -\infty.
\end{equation}
After the modes are classified, $\boldsymbol{V}$ and $\boldsymbol{\psi}$ can be reorganized as
\begin{equation}
    \boldsymbol{V} = \begin{bmatrix}\boldsymbol{V}_{+}&\boldsymbol{V}_{-}\end{bmatrix}, \quad \boldsymbol{\psi} = \begin{bmatrix}\boldsymbol{\psi}_{+}\\\boldsymbol{\psi}_{-}\end{bmatrix},
\end{equation}
where $\boldsymbol{V}_{+}$ and $\boldsymbol{V}_{-}$ contain the eigenvectors of rightgoing and leftgoing modes, respectively, and $\boldsymbol{\psi}_{+}$ and $\boldsymbol{\psi}_{-}$ contain the corresponding expansion coefficients. Using this partitioning, (\ref{Equation: linear combination1}) can be written as
\begin{equation}\label{Equation: linear combination2}
    \hat{\boldsymbol{\phi}}(x) = \sum_{i=1}^{N_+}{\boldsymbol{v}_k(x)\psi_k(x)}+\sum_{i={N_+}+1}^{N_++N_-}{\boldsymbol{v}_k(x)\psi_k(x)} = \boldsymbol{V}_{+} \boldsymbol{\psi}_{+}+\boldsymbol{V}_{-}\boldsymbol{\psi}_{-}.
\end{equation}
Here, $N_+$ and $N_{-}$ are the number of rightgoing and leftgoing modes, respectively, and $N_+ + N_- = N$. 

If (\ref{Equation: spatial marching equation}) is spatially integrated in the positive $x$ (rightgoing) direction, leftgoing modes that physically represent waves that decay as they travel in the negative $x$ direction will be wrongly captured as growing rightgoing modes, leading to exponential instability in the march. In order to make the problem well-posed and stabilize the march, the contribution of leftgoing modes, i.e. the second summation in (\ref{Equation: linear combination2}), must be eliminated. In the following section, we will discuss how to eliminate the leftgoing modes, beginning with an exact formulation followed by a computationally efficient approximation.  



\subsection{Exact one-way projection operator}

A projection operator that exactly eliminates leftgoing modes can be written in terms of the eigenvectors of $\boldsymbol{M}$ \cite{towne2016advancements, Towne2021fast}, i.e., 
\begin{equation}\label{Equation: exact filter}
    \boldsymbol{P}= \boldsymbol{V}\boldsymbol{E}\boldsymbol{V}^{-1} =\begin{bmatrix}\boldsymbol{V}_{+}&\boldsymbol{V}_{-}\end{bmatrix} \begin{bmatrix}\boldsymbol{I}_{++}& \\ &\boldsymbol{0}_{--}\end{bmatrix} \begin{bmatrix}\boldsymbol{V}_{+}&\boldsymbol{V}_{-}\end{bmatrix}^{-1},
\end{equation}
where $\boldsymbol{I}_{++} \in R^{N_+ \times N_+}$ is an identity matrix and $\boldsymbol{0}_{--} \in R^{N_- \times N_-}$ is the null matrix.  Applying this projection operator to the state vector $\hat{\boldsymbol{\phi}}$ gives
\begin{equation}\label{equation: projected state vectors, OWNS-R}
    \hat{\boldsymbol{\phi}}^{\prime} \triangleq \boldsymbol{P}\hat{\boldsymbol{\phi}} = \boldsymbol{P} \boldsymbol{V}\boldsymbol{\psi} = \boldsymbol{V}_{+} \boldsymbol{\psi}_{+}= \sum_{i=1}^{N_+}{\boldsymbol{v}_{i}\psi_i}.
\end{equation}
It is clear from~(\ref{equation: projected state vectors, OWNS-R}) that the projection operator eliminates the leftgoing modes such that the projected state vector $\hat{\boldsymbol{\phi}}^{\prime}$ contains only rightgoing modes, as desired.  Equivalently, this can be expressed as a relationship between the expansion coefficients for the projected and original states,
\begin{equation}
\label{Eq:psi_prime_eq_E_psi}
\boldsymbol{\psi}^{\prime} = \boldsymbol{E} \boldsymbol{\psi}.
\end{equation}
Since the eigenvalues of $\boldsymbol{P}$ contained within the diagonal matrix $\boldsymbol{E}$ are one for rightgoing modes and zero for leftgoing modes, the expansion coefficients for rightgoing modes are unaltered, while the expansion coefficients for leftgoing modes have been set to zero by the projection.   

In the previous OWNS-P method, this projection operator (and approximations thereof) is used to obtain a one-way evolution equation for the projected state vector $\hat{\boldsymbol{\phi}}^{\prime}$ by projecting (\ref{Equation: spatial marching equation}).  In the present work, we instead use it to project out the influence of the leftgoing modes at each step in the march as (\ref{Equation: spatial marching equation}) is integrated.  This simplifies the formulation and implementation of the method, and we will show in Section~\ref{Sec:OWNSP_compare_implement} that these two procedures are closely related.  The equations can also be integrated in the opposite, upstream direction to obtain an approximation of the leftgoing modes using the projection $\boldsymbol{I} - \boldsymbol{P}$ in place of $\boldsymbol{P}$.  

The projection operator~(\ref{Equation: exact filter}) is exact, but its formation requires the eigen-decomposition of the spatial marching operator $\boldsymbol{M}$ at every $x$ station in the spatial march.  While such an approach has been pursued by others \cite{Harris2020well}, this is prohibitively expensive, e.g., more expensive than simply solving the global problem to begin with, except in some special cases, e.g., problems that are homogeneous in the $x$ direction or that posses a small number of relevant eigenmodes.  To make the OWNS formulation useful, efficient approximations of the projection operator are needed.


\subsection{Approximate one-way projection operator}
\label{sec:approxP}

An approximate projection operator was developed as part of the OWNS-P method \cite{towne2016advancements, Towne2021fast}, as discussed in the introduction. However, the computational cost associated with building the approximate OWNS-P projection operator remains significant; it is typically around three times more costly than OWNS-O and ten to one hundred times more costly than PSE.  In this paper, we develop a new approximation of the exact projection operator whose cost (per step) approaches that of PSE while retaining the accuracy of previous OWNS methods.  The two formulations are compared in detail in Section~\ref{Comparison to OWNS-P}.

Our new approximate projection operator takes the form
\begin{equation}
\label{equation: formulation of approximate projection_P}
\boldsymbol{P}_{N_\beta}=(\boldsymbol{I}+c\boldsymbol{Z})^{-1},    
\end{equation}
where
\begin{equation}
\label{equation: formulation of approximate projection_Z}
 \boldsymbol{Z}=\prod\limits_{j=1}^{N_\beta}{(\boldsymbol{M}-i\beta_{j}^{+}\boldsymbol{I}) (\boldsymbol{M}-i\beta_{j}^{-}\boldsymbol{I})^{-1}}. 
\end{equation}
As in previous variant of OWNS \cite{Towne2015one, towne2016advancements, Towne2021fast}, we have introduced a set of recursion parameters $\left\{\beta_{j}^{+},\beta_{j}^{-}: j=1,2,...,N_{\beta} \right\}$, and $c$ is an additional parameter whose effect will be elaborated later.  The approximate projection operator $\boldsymbol{P}_{N_{\beta}}$ can be used in place of the exact projection operator $\boldsymbol{P}$ to obtain an approximation of the projected state variable
\begin{equation}
    \label{Eq:projected_state_approx}
    \hat{\boldsymbol{\phi}}^{\prime}_{N_\beta} =\boldsymbol{P}_{N_{\beta}}\hat{\boldsymbol{\phi}}.
\end{equation}

To show how this approximation works and elucidate the properties of the approximate projection, we transform equations~(\ref{equation: formulation of approximate projection_P})~-~(\ref{Eq:projected_state_approx}) into the eigenspace of $\boldsymbol{M}$.  The eigen-decomposition of $\boldsymbol{Z}$ is
\begin{equation}
\label{Eq:Z_eig_decomp}
\boldsymbol{Z} = \boldsymbol{V} \boldsymbol{F} \boldsymbol{V}^{-1},
\end{equation}
where
\begin{equation}
\label{Eq:F_def}
\boldsymbol{F} = \prod\limits_{j=1}^{N_\beta}{(\boldsymbol{\Lambda}-\beta_{j}^{+}\boldsymbol{I}) (\boldsymbol{\Lambda}-\beta_{j}^{-}\boldsymbol{I})^{-1}}    
\end{equation}
and $\boldsymbol{V}$ and $\boldsymbol{\Lambda}$ contain, respectively, the eigenvectors and eigenvalues of $\boldsymbol{M}$, as in~(\ref{Equation: eigendecomposition of M}).  This follows from the fact that each term in the product~(\ref{equation: formulation of approximate projection_Z}) has the same eigenvectors, e.g., $\boldsymbol{M}-i\beta_{j}^{+}\boldsymbol{I} = \boldsymbol{V} (i \boldsymbol{\Lambda} - i\beta_{j}^{+}\boldsymbol{I}) \boldsymbol{V}^{-1}$.


Inserting~(\ref{Eq:Z_eig_decomp}) into~(\ref{equation: formulation of approximate projection_P}) shows that the eigen-expansion of $\boldsymbol{P}_{N_{\beta}}$ is
\begin{equation}
\label{Eq:P_approx_eig_decomp}
\boldsymbol{P}_{N_{\beta}} = \boldsymbol{V} \boldsymbol{E}_{N_{\beta}} \boldsymbol{V}^{-1}
\end{equation}
with 
\begin{equation}
\label{Eq:P_approx_eigs}
\boldsymbol{E}_{N_{\beta}} = (\boldsymbol{I} + c \boldsymbol{F} )^{-1}.
\end{equation}
Comparing~(\ref{Eq:P_approx_eig_decomp}) with~(\ref{Equation: exact filter}) shows that $\boldsymbol{P}_{N_{\beta}}$ has the same eigenvectors as $\boldsymbol{P}$.  Furthermore, the exact eigenvalues $\boldsymbol{E}$ are approximated by $\boldsymbol{E}_{N_{\beta}}$, which depends on the eigenvalues of $\boldsymbol{M}$ (contained in $\boldsymbol{\Lambda}$) and the recursion parameters through $\boldsymbol{Z}$ and $\boldsymbol{F}$.


Applying the approximate projection operator to the state vector $\hat{\boldsymbol{\phi}}$ gives
\begin{equation}\label{equation: projected state vectors approximate}
    \hat{\boldsymbol{\phi}}^{\prime}_{N_{\beta}} = \boldsymbol{P}_{N_{\beta}} \hat{\boldsymbol{\phi}} = \boldsymbol{V} \boldsymbol{E}_{N_{\beta}} \boldsymbol{\psi}
\end{equation}
and the expansion coefficients for the approximate projected state are
\begin{equation}
    \label{Eq:psi_prime_approx_eq_Eapprox_psi}
    \boldsymbol{\psi}^{\prime}_{N_{\beta}} = \boldsymbol{E}_{N_{\beta}} \boldsymbol{\psi}.
\end{equation}
Comparing~(\ref{equation: projected state vectors approximate}) and~(\ref{Eq:psi_prime_approx_eq_Eapprox_psi}) with~(\ref{equation: projected state vectors, OWNS-R}) and~(\ref{Eq:psi_prime_eq_E_psi}) reveals that $\boldsymbol{E}_{N_{\beta}}$ plays the same role in the approximate projection that $\boldsymbol{E}$ does in the exact one, which is a consequence of both operators sharing the same eigenvectors.  For the approximate projection operator to faithfully mimic the exact one, the recursion parameters must be selected such that $\boldsymbol{E}_{N_{\beta}}$ converges to $\boldsymbol{E}$, i.e., that the entries of $\boldsymbol{E}_{N_{\beta}}$ converge to one and zero for rightgoing and leftgoing modes such that the corresponding expansion coefficients are unaltered and set to zero, respectively.  

To understand how the recursions accomplish this, it is helpful to define the functions
\begin{equation}
\label{Eq:F_function}
\mathcal{F}(\alpha) = \prod\limits_{j=1}^{N_{\beta}} \frac{\alpha - \beta_{j}^{+}}{\alpha - \beta_{j}^{-}}
\end{equation}
and
\begin{equation}
\label{Eq:E_function}
\mathcal{E}(\alpha) = \frac{1}{1+c \, \mathcal{F}(\alpha)}.
\end{equation}
These functions have been defined such that $\mathcal{F}(\alpha_{k})$ and $\mathcal{E}(\alpha_{k})$ return the $k$-th entries of the diagonal matrices $\boldsymbol{F}$ and $\boldsymbol{E}_{N_{\beta}}$, respectively.   Defining these functions is made possible by the fact that each eigenvalue of $\boldsymbol{P}_{N_{\beta}}$ depends only on the corresponding eigenvalue of $\boldsymbol{M}$.  As discussed later in Section~\ref{Comparison to OWNS-P}, previous OWNS variants do not exhibit this complete decoupling of each eigenmode, a property that offers several advantages.  

Recall that our goal is to choose the recursion parameters such that the eigenvalues $\boldsymbol{E}_{N_{\beta}}$ of the approximate projection matrix $\boldsymbol{P}_{N_{\beta}}$ are close to one and zero for rightgoing and leftgoing modes, respectively.  That is, we want the function $\mathcal{E}(\alpha)$ to take on these values in regions of the complex $\alpha$ plane where rightgoing and leftgoing modes reside.  This will be accomplished by choosing recursion parameters that drive the function $\mathcal{F}(\alpha)$ to zero and infinity in the vicinity of rightgoing and leftgoing modes, respectively.  Note that the function $\mathcal{F}(\alpha)$ is also defined in previous variants of OWNS, and the recursion parameters are defined to accomplish the same goals.  An important consequence of this similarity is that recursion parameters developed for previous OWNS variants can also be used for the improved formulation developed in this paper.

As described by Towne et al. \cite{Towne2015one}, a general strategy for choosing recursion parameters can be devised by considering individual terms from the product in~(\ref{Eq:F_function}),
\begin{equation}
\label{Eq:F_function_j}
\mathcal{F}_{j}(\alpha) = \frac{\alpha - \beta_{j}^{+}}{\alpha - \beta_{j}^{-}}.
\end{equation}
For a given $j$, this term will be smaller than one, and thus contribute to driving $\mathcal{E}(\alpha)$ to one, for any eigenvalue that is closer in the complex plane to $\beta_{j}^{+}$ than to $\beta_{j}^{-}$.  Conversely, $\mathcal{F}_{j}(\alpha)$ will be larger than one, and thus contribute to driving $\mathcal{E}(\alpha)$ to zero, for any eigenvalue that is closer in the complex plane to $\beta_{j}^{-}$ than to $\beta_{j}^{+}$.  Therefore, the $\beta_{j}^{+}$ and $\beta_{j}^{-}$ parameters should be placed near rightgoing and leftgoing modes, respectively, to achieve the desired properties of the projection operator.  Towne et al. \cite{Towne2015one} showed that finite sets of recursion parameters that drive $\mathcal{F}(\alpha)$ to its desired limits always exist, and effective parameters have been developed for a variety of flows, including mixing layers \citep{Towne2013improved}, subsonic and supersonic jets \citep{Towne2014continued}, and subsonic \citep{Rigas2017stability} and hypersonic \citep{kamal2020application} boundary layers.

The present OWNS formulation~(\ref{equation: formulation of approximate projection_P}) contains one additional parameter, $c$.  It is clear in~(\ref{Eq:E_function}) that large and small values of $c$ drive $\mathcal{E}(\alpha)$ toward zero and one, respectively.  In other words, large values of $c$ aid $\mathcal{F}(\alpha)$ in eliminating leftgoing modes, while small values aid $\mathcal{F}(\alpha)$ in accurately retaining rightgoing modes.  Thus, $c$ can be used to prioritize the accuracy of the method in retaining rightgoing modes or eliminating leftgoing modes.  Since both objectives are important, we recommend setting $c \sim O(1)$ in practice.


\subsection{Implementation of approximate one-way Navier-Stokes equations}

Conceptually, the approximate projection operator $\boldsymbol{P}_{N_{\beta}}$ can be applied in the same way as the exact one, i.e., by forming the operator and multiplying it with the state vector at each step in the spatial march to remove the contributions of leftgoing modes.  However, forming the approximate projection operator using~(\ref{equation: formulation of approximate projection_P}) has two disadvantages.  First, building the matrix $\boldsymbol{Z}$ via the successive multiplications in~(\ref{equation: formulation of approximate projection_Z}) lead to errors due to finite precision arithmetic symptomatic of the fact that its eigenvalues are designed to be either very large (for leftgoing modes) or very small (for rightgoing modes).  This issue becomes increasingly severe as $N_{\beta}$ increases, preventing convergence and leading to errors in the projection operator that inhibit stabilization of the march.  

Second, directly forming the approximate projection operator is computationally expensive due to the need to compute inverse matrices.  This can be overcome by solving a linear system whose solution gives the action of the approximate projection operator on a vector.  The most straightforward approach to doing so would be to define a sequence of auxiliary variables and  solve an expanded system of equations whose size scales with $N N_{\beta}$, as in previous variants of OWNS.  However, this can be avoided in the present formulation due to the comparative simplicity of the recursions.  In what follows, we show how the action of the approximate projection operator on a vector can be computed by solving a sequence of $N_{\beta}$ smaller systems of size $N$; it is this attribute of OWNS-R that enables the improved scaling of the method and brings its cost down to a level commensurate with PSE.

First, we write $\boldsymbol{Z}$ as
\begin{equation}
\boldsymbol{Z}=\boldsymbol{\Pi}^{+} (\boldsymbol{\Pi}^{-})^{-1},
\end{equation}
where 
\begin{subequations}
    \begin{align}
        &\boldsymbol{\Pi}^{+} =\prod\limits_{j=1}^{N_\beta}{(\boldsymbol{M}-i{\beta_j}^{+}\boldsymbol{I})}, \\
        &\boldsymbol{\Pi}^{-} =\prod\limits_{j=1}^{N_\beta}{(\boldsymbol{M}-i{\beta_j}^{-}\boldsymbol{I})}.\label{Eq:Pi_minus_def}
    \end{align}
\end{subequations}
Then the approximate projection matrix $\boldsymbol{P}_{N_\beta}$ can be written as
  \begin{equation}
  \label{Eq:P_approx_eq_Pm_Pstar_inv}
      \boldsymbol{P}_{N_\beta} = (\boldsymbol{I}+c\boldsymbol{Z})^{-1}=\boldsymbol{\Pi}^{-}(\boldsymbol{\Pi}^{-}+c\boldsymbol{\Pi}^{+})^{-1}={\boldsymbol{\Pi}^{-}}(\boldsymbol{\Pi}^{*})^{-1},
  \end{equation}
where
\begin{equation}
    \label{Eq:Pi_def}
    \boldsymbol{\Pi}^*=\boldsymbol{\Pi}^{-}+c\boldsymbol{\Pi}^{+}.
\end{equation}
Since both $\boldsymbol{\Pi}^+$ and $\boldsymbol{\Pi}^-$ are $N_\beta$-order polynomials of $\boldsymbol{M}$, so is $\boldsymbol{\Pi}^*$.  The matrix polynomials defining $\boldsymbol{P}_{N_{\beta}}$ in~(\ref{Eq:P_approx_eq_Pm_Pstar_inv}) can be factorized; the factorization of $\boldsymbol{\Pi}^-$ is given by its definition in~(\ref{Eq:Pi_minus_def}), while the factorization of $\boldsymbol{\Pi}^*$ must be computed from its definition in~(\ref{Eq:Pi_def}).  Critically, this matrix factorization can be computed by finding to scalar factorization
\begin{equation}
    \label{Eq:Pi_factorize_scalar}
    \left(\prod\limits_{j=1}^{N_\beta} (\alpha-\beta_j^{-}) \right) + c \left(\prod\limits_{j=1}^{N_\beta} (\alpha-\beta_j^{+}) \right) = h \prod\limits_{j=1}^{N_\beta} (\alpha-\beta_j^{*}),
\end{equation}
\sloppy{where $h$ is the coefficient of the highest-order term on the left-hand-side and the parameters $\left\{\beta_{j}^{*}: j=1,2,...,N_{\beta} \right\}$ are the roots to be determined.  Given the recursion parameters $\left\{\beta_{j}^{+}, \beta_{j}^{-}: j=1,2,...,N_{\beta} \right\}$, this factorization can be easily calculated using standard polynomial factorization functions, such as the \texttt{roots} function in MATLAB.  Pulling out the scaling factor $h$ is not theoretically necessary but helps with the numerical conditioning of the root finding procedure. } 

Using this factorization, $\boldsymbol{\Pi}^*$ can be written as
 \begin{equation}
      \boldsymbol{\Pi}^{*} = h \prod\limits_{j=1}^{N_\beta}{(\boldsymbol{M}-i\beta_{j}^{*}\boldsymbol{I})}.
  \end{equation}
Then, (\ref{Eq:P_approx_eq_Pm_Pstar_inv}) becomes
\begin{equation}\label{Eq:P_approx_eq_Pm_Pstar_inv_2}
      \boldsymbol{P}_{N_\beta} =\frac{1}{h} \prod\limits_{j=1}^{N_\beta}{(\boldsymbol{M}-i \beta_{j}^{*}\boldsymbol{I})}^{-1}
      ({\boldsymbol{M}-i\beta_{j}^{-}\boldsymbol{I}}).
\end{equation}

The form of~(\ref{Eq:P_approx_eq_Pm_Pstar_inv_2}) facilitates a sequential solution strategy for computing the action of the approximate projection operator on a vector that avoids the difficulties involved in explicitly forming the operator.  Specifically, this is accomplished by the recursions
\begin{subequations}\label{equation: implementation of OWNS-R}
    \begin{align}
        &\hat{\boldsymbol{\phi}}^{0} = \frac{1}{h}\hat{\boldsymbol{\phi}}, \\
        &(\boldsymbol{M}-i\beta_{j}^{*}\boldsymbol{I})\hat{\boldsymbol{\phi}}^{j}=(\boldsymbol{M}-i{\beta_j}^{-}\boldsymbol{I})\hat{\boldsymbol{\phi}}^{j-1}, j=1,2,...,N_{\beta}, \\
         &\hat{\boldsymbol{\phi}}^{\prime}_{N_\beta} = \hat{\boldsymbol{\phi}}^{N_\beta}.
    \end{align}
\end{subequations}
Given the state vector $\hat{\boldsymbol{\phi}}$ as input, the solution $\hat{\boldsymbol{\phi}}^{\prime}_{N_\beta}$ gives the approximate projected state vector. This can be easily verified by eliminating the auxiliary variables $\hat{\boldsymbol{\phi}}^{j}$ from (\ref{equation: implementation of OWNS-R}), yielding 
\begin{equation}
    \hat{\boldsymbol{\phi}}^{\prime}_{N_{\beta}} = \frac{1}{h} \prod\limits_{j=1}^{N_\beta}{(\boldsymbol{M}-{i \beta_j^*}\boldsymbol{I})}^{-1}
      ({\boldsymbol{M}-i{\beta_j}^{-}\boldsymbol{I}}) \, \hat{\boldsymbol{\phi}} =\boldsymbol{P}_{N_\beta} \hat{\boldsymbol{\phi}}.
\end{equation}
We emphasize that the matrices $\boldsymbol{Z}$ and $\boldsymbol{P}_{N_{\beta}}$ are never formed in practice; instead, the action of the approximate projection on a vector is obtained by solving the recursion equation~(\ref{equation: implementation of OWNS-R}).



\subsection{Computational cost scaling}
\label{sec:cost}

The primary advantage of our new OWNS formulation compared to previous OWNS variants is its improved cost scaling with the recursion order $N_{\beta}$.  To make the cost scaling explicit, recall that $N$ is the dimension of the semi-discretized state vector, i.e., the number of transverse discretization points times the number of continuous state variables, and define $N_x$ to be the number of discretization points in $x$, i.e., the number of streamwise stations in the spatial march.

First, we establish the cost scaling of OWNS-R.  Solving for $\hat{\boldsymbol{\phi}}^{j}$ in each step of the recursion~(\ref{equation: implementation of OWNS-R}) requires solving a linear system of size $N$.  The CPU and memory costs of this solution scale like $O(N^{a})$ and $O(N^{b})$, respectively, where $a$ and $b$ depend on the properties of the linear system.  For a generic dense system, $a=3$ and $b=2$.  However, sparse disctrizations such as the collocation methods used in the derivation of OWNS lead to sparse $\boldsymbol{M}$, resulting in improved scaling, often $1.5<a<2$ and $1.2<b<2$ \cite{Amestoy2019bridging}.  Computing the action of the approximate projection operator on the state vector via~(\ref{equation: implementation of OWNS-R}) requires the solution of $N_{\beta}$ linear systems, and the projection must be applied at each step in the march.  Thus, the total CPU cost of OWNS-R scales like $O(N^{a}N_{\beta}N_x)$.  Notice that the linear scaling with the recursion order $N_{\beta}$ is a consequence of the successive form of the recursion~(\ref{equation: implementation of OWNS-R}), i.e., they can be solved one at a time, beginning with the known state vector and ending with the desired projected state vector.  Likewise, since the size of each system to be solved at each step in the march is independent of $N_{\beta}$, the memory cost is independent of $N_{\beta}$ and $N_{x}$ and scales like $O(N^{b})$.

In contrast, previous variants of OWNS are formulated in terms of a coupled set of recursion equations that must be solved simultaneously.  Given its close connection with the present formulation, we use OWNS-P as an example.  Given the state vector $\hat{\boldsymbol{\phi}}$, the OWNS-P projection is defined by the coupled recursion equations \cite{towne2016advancements}
\begin{subequations}\label{equation: formulation of OWNS-P}
    \begin{align}
        &\hat{\boldsymbol{\phi}}^{-N_{\beta}}_{+}=0, \\
        & (\boldsymbol{M}-i{\beta_j}^{-}\boldsymbol{I})\hat{\boldsymbol{\phi}}^{-j}=(\boldsymbol{M}-i{\beta_j}^{+}\boldsymbol{I})\hat{\boldsymbol{\phi}}^{-j-1}, j=1,2,...,N_{\beta}-1, \\
        &(\boldsymbol{M}-i{\beta_0}^{-}\boldsymbol{I})\hat{\boldsymbol{\phi}}^{0} - (\boldsymbol{M}-i{\beta_0}^{+}\boldsymbol{I})\hat{\boldsymbol{\phi}}^{-1} =(\boldsymbol{M}-i{\beta_0}^{-}\boldsymbol{I})\hat{\boldsymbol{\phi}}, \\
        &(\boldsymbol{M}-i{\beta_j}^{+}\boldsymbol{I})\hat{\boldsymbol{\phi}}^{j}=(\boldsymbol{M}-i{\beta_j}^{-}\boldsymbol{I})\hat{\boldsymbol{\phi}}^{j+1}, j=0,1,...,N_{\beta}-1, \\
        &\hat{\boldsymbol{\phi}}^{N_{\beta}}_{-}=0.
    \end{align}
\end{subequations}
Here, the projected state is recovered from the auxiliary variables as $\hat{\boldsymbol{\phi}}^{\prime} = \hat{\boldsymbol{\phi}}^{0}$.  The plus and minus subscripts indicate taking the first $N_{+}$ and the last $N_{-}$ entries in a vector, respectively.  Notice that the known inputs $\hat{\boldsymbol{\phi}}$ and  $\hat{\boldsymbol{\phi}}^{-N_{\beta}}_{+}= \hat{\boldsymbol{\phi}}^{N_{\beta}}_{-}=0$ enter the recursions in three different places.  Because of this, the interior recursions in~(\ref{equation: formulation of OWNS-P})(b-d) cannot be solved successively, and instead the entire system must be solved simultaneously.  The size of this system is $2 N_{\beta} N$, so solving such a system at each step in the spatial march leads to an overall CPU scaling of $O(( {2 N_{\beta} N})^{a} N_x)$ and memory scaling of $O(( {2 N_{\beta} N})^{b})$.  OWNS-O leads to the same scaling, but without the factor of 2 \cite{Towne2015one, towne2016advancements}.


We can also compare these OWNS-R scalings with those obtained for other methods.  For global linear methods, a linear system of size $N{N_x}$ is solved, leading to CPU and memory cost scalings of $O((NN_x)^{a})$ and $O((NN_x)^{b})$, respectively.  The relative CPU cost of a global method compared to OWNS-R is thus $O(N_x^{a-1} / N_{\beta})$.  Since $N_x >> N_{\beta}$, this can lead to substantial speedups.  The relative memory cost of a global method compared to OWNS-R is $O(N_x^{b})$, so the memory savings achieved by OWNS-R are always large.  An exact formulation of any of the OWNS methods requires the eigendecomposition of $\boldsymbol{M}$ at each step in the march, which scales like $O(N^3 N_x)$.  Therefore, the relative CPU cost compared to OWNS-R is $O(N^{3-a}/N_{\beta})$.  Since $N >> N_{\beta}$, the approximate OWNS-R formulation is typically orders of magnitude faster than an exact formulation.  

Finally, in PSE a linear system of size $N$ must be solved $N_{it}$ times at each step in the march, leading to CPU and memory cost scalings of $O(N^{a} N_x N_{it})$ and $O(N^{b})$. Here $N_{it}$ is the number of iterations required to satisfy a nonlinear constraint that is part of the PSE formulation. The relative CPU cost compared to OWNS-R is thus $O(N_{it}/N_{\beta})$.  Typically $N_{it}$ is slightly smaller but similar in size to $N_{\beta}$.  This suggests that OWNS-R can approach the speed of PSE while capturing all rightgoing modes rather than just one, and this will be verified using several example problems in Section~\ref{sec:examples}.  Furthermore, this improved OWNS-R solution requires no additional memory compared to PSE.

\setlength{\tabcolsep}{24pt}
\begin{table}
\centering
\caption{\small Cost Scaling: $1<a<3$ and $1<b<2$, with typically values $a\approx 1.5$ and $b\approx 1.2$.}
\label{tab:theoretical_scaling}
\renewcommand\arraystretch{1.33}
\begin{tabular}{lll}
\hline
\textbf{Method} & \textbf{CPU}                & \textbf{Memory}        \\ 
\hline
Global          & $(N N_{x})^{a}$             & $(N N_{x})^{b}$        \\ 

PSE             & $N^{a} N_{it} N_{x}$        & $N^{b}$                \\ 

OWNS-O          & $(N N_{\beta})^{a} N_{x}$   & $(N N_{\beta})^{b}$    \\ 

OWNS-P          & $(2 N N_{\beta})^{a} N_{x}$ & $(2 N N_{\beta})^{b}$  \\ 

OWNS-R          & $N^{a} N_{\beta} N_{x}$     & $N^{b}$       \\
\hline
\end{tabular}
\end{table}

\subsection{Error analysis}

Next, we derive an expression for the error produced by the approximate projection operator relative to the exact one.  Specifically, we define the error as the norm of the difference between the approximate and true projected state, normalized by the norm of the un-projected state, 
\begin{equation}
\label{Eq:error_def_phi}
error =  \frac{{\Vert\hat{\boldsymbol{\phi}}^{\prime}_{N_{\beta}} - \hat{\boldsymbol{\phi}}^{\prime} \Vert}_2}{{\Vert\hat{\boldsymbol{\phi}}\Vert}_2} \leq {\Vert \boldsymbol{P}_{N_\beta} - \boldsymbol{P}\Vert}_2.
\end{equation}
The inequality follows from the identify ${\Vert\hat{\boldsymbol{\phi}}^{\prime}_{N_{\beta}} - \hat{\boldsymbol{\phi}}^{\prime} \Vert}_2 = {\Vert {\boldsymbol{P}}_{N_{\beta}}\hat{\boldsymbol{\phi}} - {\boldsymbol{P}}  \hat{\boldsymbol{\phi}} \Vert}_2 \leq {\Vert \boldsymbol{P}-\boldsymbol{P}_{N_\beta}\Vert}_2 \, {\Vert \hat{\boldsymbol{\phi}} \Vert}_{2}$.

Using~(\ref{Equation: exact filter}) to replace $\boldsymbol{P}$ and~(\ref{Eq:P_approx_eig_decomp}) and~(\ref{Eq:P_approx_eigs}) to replace $\boldsymbol{P}_{N_{\beta}}$, the difference between the approximate and exact projection operators appearing in (\ref{Eq:error_def_phi}) becomes
\begin{equation}
\label{Eq:error_P_diff_1}
    \boldsymbol{P}_{N_\beta} - \boldsymbol{P} = { \boldsymbol{V}\begin{bmatrix}{(\boldsymbol{I}_{++}+c\boldsymbol{F}_{++})^{-1}- \boldsymbol{I}_{++}}&O\\O&{(\boldsymbol{I}_{--}+c\boldsymbol{F}_{--}})^{-1}\end{bmatrix} \boldsymbol{V}^{-1}},
\end{equation}
where $\boldsymbol{F}_{++} \in \mathbb{C}^{N_+ \times N_+}$ and $\boldsymbol{F}_{--} \in \mathbb{C}^{N_- \times N_-}$ are a diagonal matrices containing the entries of $\boldsymbol{F}$ related to rightgoing and leftgoing modes, respectively, and $\boldsymbol{I}_{++}$ and $\boldsymbol{I}_{--}$ are identity matrices of appropriate dimension.  

Since our recursion parameters have been chosen to drive the function $\mathcal{F}(\alpha)$ toward zero for rightgoing modes and infinity for leftgoing modes, we assume that ${\Vert \boldsymbol{F}_{++}\Vert}_2 \ll {\Vert \boldsymbol{I}_{++}\Vert}_2,  {\Vert \boldsymbol{F}_{--}\Vert}_2 \gg{\Vert \boldsymbol{I}_{--}\Vert}_2 $ for sufficiently large $N_{\beta}$.  This permits the first-order series approximations
\begin{subequations}
    \begin{align}
        &(\boldsymbol{I}_{++}+c\boldsymbol{F}_{++})^{-1}\approx{\boldsymbol{I}_{++}-c\boldsymbol{F}_{++}}, \\
        &(\boldsymbol{I}_{--}+c\boldsymbol{F}_{--})^{-1}\approx (c\boldsymbol{F}_{--})^{-1}.
    \end{align}
\end{subequations}
Using these approximations, (\ref{Eq:error_P_diff_1}) takes the simplified form
\begin{equation}
    \boldsymbol{P}_{N_\beta} - \boldsymbol{P} \approx{ \boldsymbol{V}\begin{bmatrix}{({\boldsymbol{I}_{++}-c\boldsymbol{F}_{++}} - \boldsymbol{I}_{++})}&O\\O&{(c\boldsymbol{F}_{--})^{-1}}\end{bmatrix} \boldsymbol{V}^{-1}}={ \boldsymbol{V}\begin{bmatrix}{-c\boldsymbol{F}_{++}}&O\\O&(c\boldsymbol{F}_{--})^{-1}\end{bmatrix} \boldsymbol{V}^{-1}}.
\end{equation}
Using this result in~(\ref{Eq:error_def_phi}), the projection error can be expressed as
\begin{equation}\label{error analysis for correction2}
    error \leq {\Vert \boldsymbol{P}_{N_{\beta}} - \boldsymbol{P}\Vert}_2 \leq {{\Vert{\boldsymbol{V}}\Vert}_2} \, {{\Vert{\boldsymbol{V}^{-1}}\Vert}_2} \, max({c\Vert{\boldsymbol{F}_{++}}\Vert}_2,{{c}^{-1}\Vert \boldsymbol{F}_{--}^{-1}\Vert}_2).
\end{equation}
Equation~(\ref{error analysis for correction2}) shows that the maximum error is influenced by three factors.  The first is the value of ${{\Vert{\boldsymbol{V}}\Vert}_2} {{\Vert{\boldsymbol{V}^{-1}}\Vert}_2}$, which is the condition number of the eigenvector matrix $\boldsymbol{V}$.  Fluid systems often exhibit non-normal operators leading to the potential of large condition numbers.  The second factor is the maximum amplitude of $\boldsymbol{F}_{++}$ and $\boldsymbol{F}_{--}^{-1}$, which depends on the number and locations of the recursion parameters and thus can be made arbitrarily small by increasing the order of the recursions.  Finally, as discussed previously, the parameter $c$ can be used to balance the error in retaining rightgoing modes and eliminating leftgoing modes.  

The preceding analysis provides an upper bound for the worst-case error.  More useful expressions can be obtained by considering the error of individual modes.  From~(\ref{Equation: linear combination1}), the contribution of the $k$-th mode to the state vector is $\boldsymbol{v}_{k} \psi_{k}$.  The error produced by projecting this component of the solution with the approximate projection rather than the exact one is
\begin{equation}
\label{Eq:error_def_v}
error_{k} = \frac{{\Vert {\boldsymbol{P}}_{N_{\beta}}\boldsymbol{v}_{k} \psi_{k} - {\boldsymbol{P}}  \boldsymbol{v}_{k} \psi_{k} \Vert}_2}{{\Vert\boldsymbol{v}_{k} \psi_{k} \Vert}_2}.
\end{equation}
Using the eigen-expansions for $\boldsymbol{P}_{N_{\beta}}$ and $\boldsymbol{P}$ given in~(\ref{Eq:P_approx_eig_decomp}) and~(\ref{Equation: exact filter}), respectively, the difference in the numerator of~(\ref{Eq:error_def_v}) can be written as
\begin{equation}
    \label{Eq:error_P_diff_v}
    \boldsymbol{V} \left( \boldsymbol{E}_{N_\beta} - \boldsymbol{E} \right) \boldsymbol{V}^{-1} \boldsymbol{v}_{k}.
\end{equation}
Since the product $\boldsymbol{V}^{-1} \boldsymbol{v}_{k}$ yields the $k$-th row of the identity matrix, (\ref{Eq:error_P_diff_v}) simplifies to $\boldsymbol{v}_{k}$ times the difference between the $k$-th eigenvalues of $\boldsymbol{P}_{N_{\beta}}$ and $\boldsymbol{P}$ located in the diagonal matrices $\boldsymbol{E}_{N_\beta}$ and $\boldsymbol{E}$.  Recall that the $k$-th eigenvalue of $\boldsymbol{P}_{N_{\beta}}$ is given by $\mathcal{E}(\alpha_k)$, while the eigenvalues of $\boldsymbol{P}$ are one for leftgoing modes and zero for rightgoing modes.  Using this result in~(\ref{Eq:error_def_v}) gives an explicit expression for the error for the $k$-th mode,
\begin{equation}
\label{Eq:error_v_final}
error_{k} = \frac{{\Vert {\boldsymbol{P}}_{N_{\beta}}\boldsymbol{v}_{k} \psi_{k} - {\boldsymbol{P}}  \boldsymbol{v}_{k} \psi_{k} \Vert}_2}{{\Vert\boldsymbol{v}_{k} \psi_{k} \Vert}_2} = \left\{
\begin{aligned}
&\mathcal{E}(\alpha_{k})-1 \quad &&\text{for rightgoing modes}, \\
&\mathcal{E}(\alpha_{k}) \quad &&\text{for leftgoing modes}.
\end{aligned}
\right.
\end{equation}
The recursion parameters are selected to force $\mathcal{E}(\alpha_{k})$ toward one and zero for rightgoing and leftgoing modes, respectively, driving the error toward zero.

Since $\mathcal{E}(\alpha)$ is a function only of the recursion parameters, the error for any possible eigenvalue $\alpha$ can be assessed \emph{a priori} with no need for knowledge of the associated eigenvector or other modes of the system.  This result is a consequence of the exact and approximate projection operators sharing the same eigenvectors.  As a result, the error does not depend on the condition number of the eigenvector matrix $\boldsymbol{V}$, despite its appearance in the upper-bound error estimate in~(\ref{equation: solution error}).  We emphasize that exact error expressions in terms of eigenvalues alone cannot be derived for previous OWNS variants; the error produced by OWNS-O and OWNS-P does depend on the eigenvectors of the system and can be impacted by their non-normality.

%


\section{Comparison to OWNS-P}
\label{Comparison to OWNS-P}

As mentioned already, the present OWNS-R formulation and the previous OWNS-P method are closely related in that they both produce an approximation of the same exact projection operator $\boldsymbol{P}$.  The cost savings achieved by OWNS-R compared to OWNS-P were highlighted in Section~\ref{sec:cost}.  In this section, we compare two further differences between these methods: the nature of the approximation of the exact projection operator delivered by each method and the way in which the projection operator is applied to achieve a stable spatial march.


\subsection{Approximation of the projection operator}

OWNS-R and OWNS-P produce two different approximations of the same exact projection operator $\boldsymbol{P}$. As a point of comparison, recall that the exact projection operator can be written in the form 
\begin{equation}\label{Equation: exact filter2}
    \boldsymbol{P}= \boldsymbol{V} \begin{bmatrix}\boldsymbol{I}_{++}& \\ &\boldsymbol{0}_{--}\end{bmatrix} \boldsymbol{V}^{-1}.
\end{equation}
We showed in Section~\ref{sec:approxP} that the OWNS-R approximation of the projection operator can be written in terms of its eigen-decomposition as
\begin{equation}
\label{Eq:P_approx_eig_decomp_2}
\boldsymbol{P}_{N_{\beta}} = \boldsymbol{V} (\boldsymbol{I} + c \boldsymbol{F} )^{-1} \boldsymbol{V}^{-1}.
\end{equation}
Finally, Towne et al.\cite{Towne2021fast} showed that the projection operator implicitly defined by the OWNS-P recursion in~(\ref{equation: formulation of OWNS-P}) can be written in terms of its eigen-decomposition as
\begin{equation}\label{Eq:OWNSP_decomp}
    \tilde{\boldsymbol{P}}_{N_{\beta}} =  \tilde{\boldsymbol{V}} \begin{bmatrix}\boldsymbol{I}_{++}& \\ &\boldsymbol{0}_{--}\end{bmatrix} \tilde{\boldsymbol{V}}^{-1},
\end{equation}
where
\begin{equation}
\label{Eq:OWEP_P_approx}
\tilde{\boldsymbol{V}} = \left[ \begin{array}{cc} \boldsymbol{V}_{+} & \boldsymbol{V}_{-} \end{array} \right] \left[\begin{array}{cc} \boldsymbol{I}_{++} & -\boldsymbol{R}_{+-} \\ -\boldsymbol{R}_{-+} & \boldsymbol{I}_{--} \end{array}\right]^{-1}
\end{equation}
is a matrix containing the eigenvectors of $\tilde{\boldsymbol{P}}_{N_{\beta}}$ and $\boldsymbol{R}_{+-}$ and $\boldsymbol{R}_{-+}$ are matrices that converge toward zero as the order of the recursions increases.

Comparing~(\ref{Eq:P_approx_eig_decomp_2}) and~(\ref{Eq:OWNSP_decomp}) with~(\ref{Equation: exact filter2}) reveals a fundamental, symmetric difference between the two approximations: the OWNS-P operator $\tilde{\boldsymbol{P}}_{N_{\beta}}$ has exact eigenvalues but approximate eigenvectors, while the OWNS-R operator  $\boldsymbol{P}_{N_{\beta}}$ has approximate eigenvalues but exact eigenvectors. Both approximations converge toward the same exact projection operator $\boldsymbol{P}$ as the order of the recursions increases: the eigenvalues of ${\boldsymbol{P}}_{N_{\beta}}$ converge toward their exact values as $\boldsymbol{F}$ tends toward zero for rightgoing modes and infinity for leftgoing modes, while the eigenvectors of $\tilde{\boldsymbol{P}}_{N_{\beta}}$ converge toward $\boldsymbol{V}$ as $\boldsymbol{R}_{+-}$ and $\boldsymbol{R}_{-+}$ tend toward zero.


\subsection{Stabilization of the spatial march} 
\label{Sec:OWNSP_compare_implement}

Next, we compare the way in which the projection operator is used to stabilize the spatial march in OWNS-P and OWNS-R.  For simplicity, we assume that the recursions are sufficiently converged such that the exact projection operator $\boldsymbol{P}$ can be used in place of the approximate projection operators for both methods.

In OWNS-P, the projection operator is used to derive an evolution equation for the projected state variable $\hat{\boldsymbol{\phi}}^{\prime}$.  Multiplying~(\ref{Equation: spatial marching equation}) by $\boldsymbol{P}$, using the fact that the projection operator commutes with $\boldsymbol{M}$ and satisfies the condition $\boldsymbol{PP}=\boldsymbol{P}$, and negelcting $x$ derivatives of $\boldsymbol{P}$ yields an evolution equation for the projected state in continuous $x$ space,
\begin{equation} \label{Equation: projected spatial marching equation2}
    \frac{\mathrm{d}\hat{\boldsymbol{\phi}}^{\prime}}{\mathrm{d}x}=\boldsymbol{PM}\hat{\boldsymbol{\phi}}^{\prime} +\boldsymbol{P} \hat{\boldsymbol{g}}.
\end{equation}
Since the projection operator removes all leftgoing modes from the evolution operator $\boldsymbol{M}$, (\ref{Equation: projected spatial marching equation2}) can be stably integrated in the positive $x$ direction.  While theoretically sound, Towne \cite{towne2016advancements} observed that this approach can lead to an accumulation of energy in leftgoing modes due to numerical errors.  These leftgoing waves can be eliminated by applying the projection operator again after each step in the march.

In our OWNS-R formulation, we forgo the use of the $x$-continuous one-way equation~(\ref{Equation: projected spatial marching equation2}) and instead rely entirely on projecting the state variable after each step in the integration of~(\ref{Equation: spatial marching equation}) to stabilize the march.  In what follows, we show that these two approaches are closely related.  The derivation of this point is dependent on the integration scheme.  For brevity, we focus our attention on the prototypical explicit and implicit integration schemes -- the explicit and implicit Euler methods -- but we note that similar results and conclusions can be obtained for broad classes of linear multistep and Runge-Kutta methods. 

For an explicit Euler integration of the one-way evolution equation~(\ref{Equation: projected spatial marching equation2}), the projected state variable is advanced from $x_n$ to $x_{n+1} = x_n + \Delta x$ as
\begin{equation}\label{Equation: step equation for projected spatial marching equaiton using explicit Euler}
    \hat{\boldsymbol{\phi}}^{\prime}_{n+1} = (\boldsymbol{I}+\Delta x \boldsymbol{P}_n \boldsymbol{M}_n)\hat{\boldsymbol{\phi}}^{\prime}_n + \Delta x \boldsymbol{P}_n \hat{\boldsymbol{g}}_n,
\end{equation}
where the $n$ and $n+1$ subscripts indicate where the operators are evaluated, e.g., $\boldsymbol{P}_n$ is the projection operator at $x = x_n$.  Subsequent application of the projection operator $P_{n+1}$, as described above, gives the total OWNS-P step,
\begin{equation}\label{Eq:OWNSP_step}
    \hat{\boldsymbol{\phi}}^{\prime}_{n+1} = \boldsymbol{P}_{n+1}(\boldsymbol{I}+\Delta x \boldsymbol{P}_n \boldsymbol{M}_n)\hat{\boldsymbol{\phi}}^{\prime}_n + \Delta x \boldsymbol{P}_{n+1} \boldsymbol{P}_n \hat{\boldsymbol{g}}_n.
\end{equation}

The OWNS-R explicit Euler step is obtained by first applying explicit Euler integration of the (un-projected) evolution equation~(\ref{Equation: spatial marching equation}),
\begin{equation}\label{Equation: step equation for spatial marching equaiton using explicit Euler}
    \hat{\boldsymbol{\phi}}_{n+1} = (\boldsymbol{I}+\Delta x \boldsymbol{M}_n)\hat{\boldsymbol{\phi}}^{\prime}_n + \Delta x \hat{\boldsymbol{g}}_n,
\end{equation}
and subsequently applying the projection operator $\boldsymbol{P}_{n+1}$ to the solution, yielding
\begin{equation}\label{Eq:OWNSA_step}
    \hat{\boldsymbol{\phi}}^{\prime}_{n+1} = \boldsymbol{P}_{n+1}(\boldsymbol{I}+\Delta x \boldsymbol{M}_n)\hat{\boldsymbol{\phi}}^{\prime}_n + \Delta x \boldsymbol{P}_{n+1} \hat{\boldsymbol{g}}_n.
\end{equation}
In~(\ref{Equation: step equation for spatial marching equaiton using explicit Euler}) and~(\ref{Eq:OWNSA_step}), we used $\hat{\boldsymbol{\phi}}^{\prime}_{n}$ instead of $\hat{\boldsymbol{\phi}}_{n}$ to represent the previous state because the state vector at $x_n$ will have already been projected using $\boldsymbol{P}_n$ as part of the previous step in the march.

Comparing the OWNS-P step in~(\ref{Eq:OWNSP_step}) and the OWNS-R step in~(\ref{Eq:OWNSA_step}) revels two apparent  differences.  The first is that the $\hat{\boldsymbol{\phi}}^{\prime}_{n}$ term contains an extra instance of $\boldsymbol{P}_{n}$.  However, this $\boldsymbol{P}_{n}$ has no impact on the solution.  To see this, recall that $\boldsymbol{P}_{n}$ and $\boldsymbol{M}_{n}$ commute and that applying $\boldsymbol{P}_{n}$ to $\hat{\boldsymbol{\phi}}^{\prime}_{n}$ has no impact since it has already been projected during the previous step.  Thus, $\boldsymbol{P}_n \boldsymbol{M}_n \hat{\boldsymbol{\phi}}^{\prime}_n = \boldsymbol{M}_n \boldsymbol{P}_n \hat{\boldsymbol{\phi}}^{\prime}_n = \boldsymbol{M}_n \hat{\boldsymbol{\phi}}^{\prime}_n$, so the extra $\boldsymbol{P}_n$ is redundant and falls out of the equation.  The second difference is that the forcing $\hat{\boldsymbol{g}}_n$ is projected by $\boldsymbol{P}_{n+1} \boldsymbol{P}_n$ rather than just $\boldsymbol{P}_{n+1}$ as part of the OWNS-P step.  However, as long as the flow is slowly varying in $x$, which was already assumed in the derivation of the one-way equation and is necessary for a spatial marching equation to be appropriate in the first place, $\boldsymbol{P}_{n+1} \approx \boldsymbol{P}_n$, so the extra $\boldsymbol{P}_n$ will have little impact on the projected forcing.  Overall, then, using the original evolution equation~(\ref{Equation: spatial marching equation}) rather than the formal one-way equation~(\ref{Equation: projected spatial marching equation2}) within the OWNS-R framework along with explicit Euler integration has negligible impact on the solution.

Next, we repeat the same analysis for implicit Euler integration.  An implicit Euler step of the OWNS-P one-way evolution equation~(\ref{Equation: projected spatial marching equation2}) followed by application of $\boldsymbol{P}_{n+1}$ yields
\begin{equation}\label{Eq:OWNSP_step_IE}
\hat{\boldsymbol{\phi}}^{\prime}_{n+1} = \boldsymbol{P}_{n+1} (\boldsymbol{I}-\Delta x \boldsymbol{P}_{n+1} \boldsymbol{M}_{n+1})^{-1}(\hat{\boldsymbol{\phi}}^{\prime}_{n} + \Delta x \boldsymbol{P}_{n+1} \hat{\boldsymbol{g}}_{n+1}).
\end{equation}
Similarly, an implicit Euler step of the original evolution equation~(\ref{Equation: spatial marching equation}) followed by application of $\boldsymbol{P}_{n+1}$ yields
\begin{equation}\label{Eq:OWNSA_step_IE}
\hat{\boldsymbol{\phi}}^{\prime}_{n+1} = \boldsymbol{P}_{n+1} (\boldsymbol{I}-\Delta x \boldsymbol{M}_{n+1})^{-1}(\hat{\boldsymbol{\phi}}^{\prime}_{n} + \Delta x \hat{\boldsymbol{g}}_{n+1}).
\end{equation}
Comparing these results again reveals two apparent differences: the inverse term and the forcing term each contain an additional instance of $\boldsymbol{P}_{n+1}$ in~(\ref{Eq:OWNSP_step_IE}).  Neither impacts the solution.  Using that $\boldsymbol{P}_{n+1}$ is a projection and commutes with $\boldsymbol{M}_{n+1}$, it can be shown that 
$\boldsymbol{P}_{n+1} (\boldsymbol{I}-\Delta x \boldsymbol{P}_{n+1} \boldsymbol{M}_{n+1})^{-1} = \boldsymbol{P}_{n+1} (\boldsymbol{I}-\Delta x \boldsymbol{M}_{n+1})^{-1} $ and $\boldsymbol{P}_{n+1} (\boldsymbol{I}-\Delta x \boldsymbol{M}_{n+1})^{-1} =  \boldsymbol{P}_{n+1} (\boldsymbol{I}-\Delta x  \boldsymbol{M}_{n+1})^{-1} \boldsymbol{P}_{n+1}$. The first of these results show that the terms involving $\hat{\boldsymbol{\phi}}^{\prime}_{n}$ are equivalent in~(\ref{Eq:OWNSP_step_IE}) and~(\ref{Eq:OWNSA_step_IE}) and the second result shows that the terms involving $\hat{\boldsymbol{g}}_{n+1}$ are equivalent.  Therefore, using the original evolution equation~(\ref{Equation: spatial marching equation}) rather than the formal one-way equation~(\ref{Equation: projected spatial marching equation2}) within the OWNS-R framework along with implicit Euler integration has no impact on the solution.


\section{Example applications}
\label{sec:examples}
 
In this section, the OWNS-R method is applied to three example problems: a dipole forcing in a quiescent fluid, a Mach 0.9 turbulent jet, and a supersonic boundary layer.  For each case, the result and computational cost are compared to other methods to show the accuracy and efficiency of our new formation of OWNS. All simulations are performed on a laptop with an Intel Core i7-8750H@2.2GHz processor and 32 GB of RAM.  In all cases, recursion parameters are selected using the recipe outlined by Towne et al. \cite{Towne2015one} and the parameter $c$ is set to one.  The linearized Navier-Stokes equations are discretized in transverse directions using fourth-order central finite differences with summation-by-parts boundary closure \citep{Mattsson2004summation}. Far-field radiation boundary conditions are enforced at free transverse boundaries using a super-grid damping layer \citep{Appelo2009high} truncated by Thompson characteristic conditions \citep{Thompson1987time}. Unless otherwise noted, the spatial march in $x$ is performed using a second-order backward difference formula \cite{Hairer1996solving} for all OWNS variants, while PSE employs implicit Euler integration by construction.  Additional problem-specific details of the numerical methods are reported in the appropriate subsections that follow.


\subsection{Dipole acoustic waves in a quiescent fluid}

First, we consider propagation of acoustic waves generated by a two-dimensional dipole forcing in a quiescent, inviscid fluid. Its relative simplicity along with the ability to compute an exact OWNS solution (due to spatial homogeneity) makes this a useful case for illustrating the properties and performance of the new OWNS-R formulation.  The dipole is placed at the origin of the computational domain, i.e., $x = 0, y = 0$. The transverse domain, not including the damping layer, extends from $-5$ to $5$ and is discretized using 201 grid points. In the $x$ direction, the linearized equations are integrated from $-5$ to $5$ with a uniform step size of $\Delta x = 0.05$.  A zero initial disturbance is specified at the domain inlet and the response is generated entirely by the right-hand-side dipole forcing term.  The dipole oscillates at angular frequency $\omega = 2\pi$ such that the wavelength of the emitted acoustic waves is equal to one.  


The eigenvalues of the approximate projection operator, i.e., $\boldsymbol{E}_{N_\beta}$, are shown in Fig. \ref{Waveforcing, correction2, filter} for different numbers of recursion parameters, $N_\beta = 5,7,10,15$.  Blue and red symbols indicate rightgoing and leftgoing modes, respectively.  To remove leftgoing modes while accurately retaining rightgoing modes, the associated eigenvalues should take on values of $(0,0)$ and $(1,0)$, respectively.  When too fewer recursion parameters are used, e.g., $N_{\beta} = 5$ shown in Fig. \ref{Waveforcing, correction2, filter}(a), the eigenvalues of the projection operator for the leftgoing modes are clustered around $(0,0)$ but still deviate significantly from this value, and an analogous statement can be made about the rightgoing modes. In particular, the error in the leftgoing modes for $N_{\beta} = 5$ is large enough to prevent a stable spatial march.  With increasing $N_\beta$, the eigenvalues of the approximate projection operator converge to $(0,0)$ and $(1,0)$ for leftgoing and rightgoing modes, respectively. As a result, the leftgoing modes are eliminated and the rightgoing modes are accurately retained.

\begin{figure}[t]
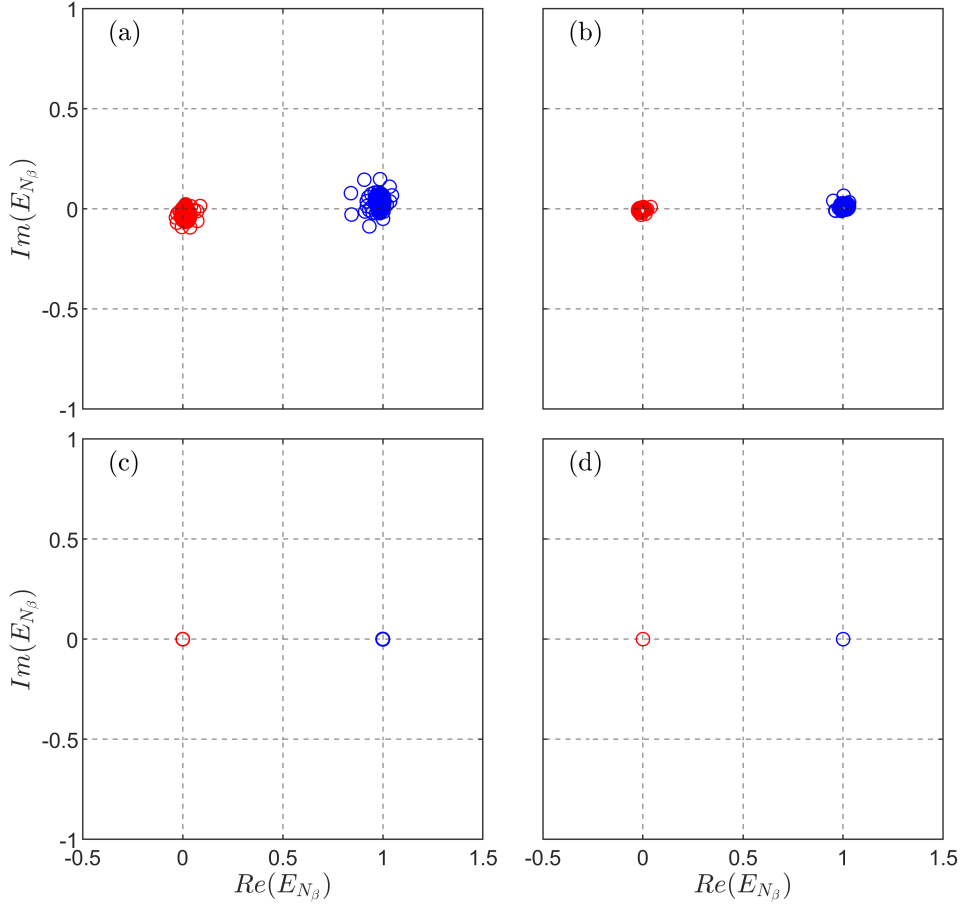

     \centering
     \psfragfig!{Picture/Final/WF_Comparison_F}
     \caption{The eigenvalues of the approximate projection matrix $\boldsymbol{P}_{N_\beta}$ for: (a) $N_{\beta}$ = 5; (b) $N_{\beta}$ = 7; (c) $N_{\beta}$ = 10; (d) $N_{\beta}$ = 15.  Symbols:  (\lowersym{\textcolor{blue}{\bf\SmallCircle}}) rightgoing modes; (\lowersym{\textcolor{red}{\bf\SmallCircle}}) leftgoing modes.  As desired, the eigenvalues converge to (0,0) and (1,0) for leftgoing and rightgoing modes, respectively, as the order of the recursions increases.} 
     \label{Waveforcing, correction2, filter}
\end{figure}

\begin{figure}[!ht]
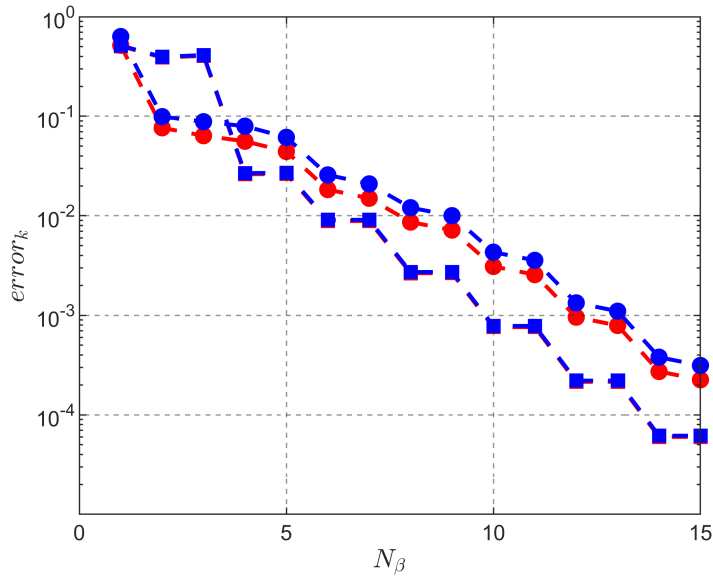

     \centering
     \psfragfig!{Picture/Final/WF_E_plot2}
     \caption{Convergence of the eigenvalues of the approximate projection matrix $\boldsymbol{P}_{N_\beta}$ as a function of the recursion order $N_{\beta}$ for: (\lowersym{\textcolor{blue}{\bf\FilledSmallCircle}}) $\alpha$ = 1.5+6.1 i; (\lowersym{\textcolor{red}{\bf\FilledSmallCircle}}) $\alpha$ = -1.5-6.1 i; (\lowersym{\textcolor{blue}{\bf\FilledSmallSquare}}) $\alpha$ = 26.8 i; (\lowersym{\textcolor{red}{\bf\FilledSmallSquare}}) $\alpha$ = -26.8 i. }
     \label{fig:dipole_E_eig_convergence}
\end{figure}

Fig.~\ref{fig:dipole_E_eig_convergence} demonstrates the quantitative convergence of the approximate projection operator by plotting $\mathcal{E}(\alpha_{k})$ and $\mathcal{E}(\alpha_{k}) - 1$, as specified in~(\ref{Eq:error_def_v}), for two leftgoing modes and rightgoing modes, i.e., $\alpha = 1.5 \pm 6.1 i$ and $\alpha = \pm 26.8i$.  The convergence is roughly exponential, as expected \cite{Towne2015one}.  

Fig.~\ref{Waveforcing, correction2, PM} provides further insights into the convergence of the approximate projection operator.  The eigenvalues of the spatial evolution operator $\boldsymbol{M}$ are indicated with pink (x) symbols in the complex $\alpha$ plane.  Eigenvalues in the upper-right and lower-left quadrants correspond to rightgoing and leftgoing modes, respectively \cite{Towne2015one}.  In each sub-figure, the green and back squares indicate the locations of the $\beta^{+}_{j}$ and $\beta^{-}_{j}$ recursion parameters, respectively, for the same four values of $N_{\beta}$ as in Fig.~\ref{Waveforcing, correction2, filter}.  As described in Section~\ref{sec:approxP}, these parameters are located near rightgoing and leftgoing eigenvalues, respectively.  The background contour levels show the magnitude of the function $\mathcal{E}(\alpha)$ for each $N_{\beta}$.  It is apparent that $\mathcal{E}(\alpha)$ is converging toward its desired limits (1 and 0) in regions occupied by rightgoing and leftgoing eigenvalues, respectively, and that the cutoff between these limiting values is becoming increasingly sharp.  Note that the exact locations of the eigenvalues of $\boldsymbol{M}$ are not needed to define the recursion parameters (they are based on an analytical approximation of their locations \cite{Towne2015one}) or to estimate the the magnitude of the eigenvalues of the approximate projection operator using the function $\mathcal{E}(\alpha)$ and thus the projection error defined in~(\ref{Eq:error_v_final}).   

Additionally, the blue and red circles in Fig.~\ref{Waveforcing, correction2, PM} show the eigenvalues of the projected spatial marching operator $\boldsymbol{P}_{N_\beta}\boldsymbol{M}$.  While our OWNS-R methodology does not directly use this projected operator, instead applying the projection directly to the state vector at each step in the march, we showed in Section~\ref{Comparison to OWNS-P} that using a projected spatial evolution operator is consistent with our approach.  Additionally, previous OWNS variants have been traditionally understood in terms of these modified eigenvalues.  For both of these reasons, considering them here is helpful for understanding  the error introduced by the approximate projection matrix. Ideally, the rightgoing eigenvalues of the projected evolution operator $\boldsymbol{P}_{N_\beta}\boldsymbol{M}$ will match the corresponding modes of $\boldsymbol{M}$, while the leftgoing modes will be set to zero.  Large deviations from this ideal behavior are observed for low values of $N_{\beta}$, but the desired results are obtained with increasing accuracy as $N_{\beta}$ is increased.

\begin{figure}[!t]
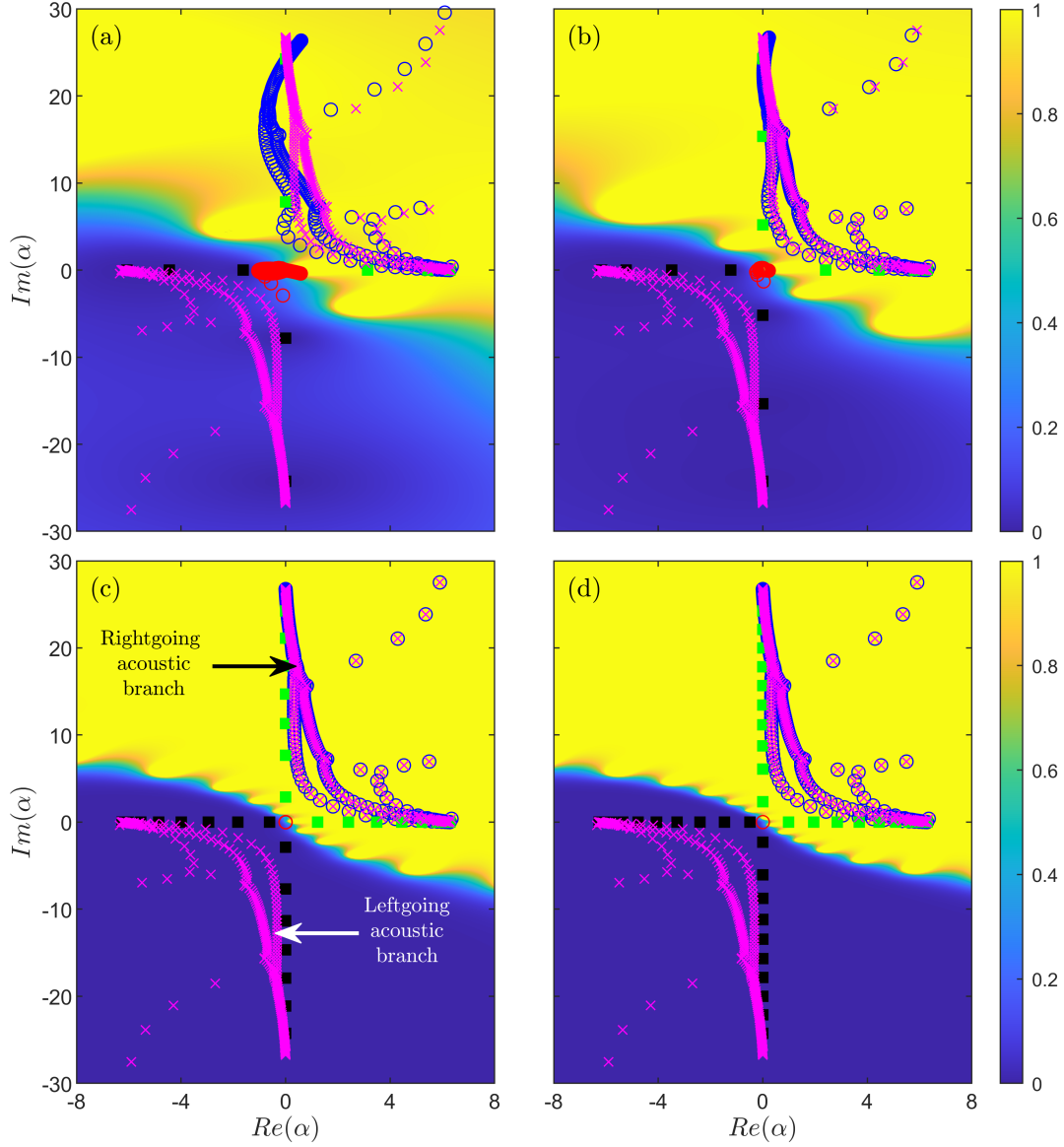

     \centering
     \psfragfig!{Picture/Final/WF_Comparison_PM}
     \caption{Eigen-space results for the dipole test case for: (a) $N_{\beta}$ = 5; (b) $N_{\beta}$ = 7; (c) $N_{\beta}$ = 10; (d) $N_{\beta}$ = 15.  Symbols: (\lowersym{\textcolor{magenta}{\bf\SmallCross}}) eigenvalues of $\boldsymbol{M}$; (\lowersym{\textcolor{blue}{\bf\SmallCircle}}) rightgoing eigenvalues of $\boldsymbol{P}_{N_\beta}\boldsymbol{M}$; (\lowersym{\textcolor{red}{\bf\SmallCircle}}) leftgoing eigenvalues of $\boldsymbol{P}_{N_\beta}\boldsymbol{M}$; (\lowersym{\textcolor{green}{\bf\FilledSmallSquare}}) ${\beta_j}^{+}$; (\lowersym{\textcolor{black}{\bf\FilledSmallSquare}}) ${\beta_j}^{-}$. The contours show the magnitude of $\mathcal{E}(\alpha)$, which converges toward 1 and 0 in the vicinity of rightgoing and leftgoing modes, respectively.}
     \label{Waveforcing, correction2, PM}
\end{figure}

\begin{figure}[!t]
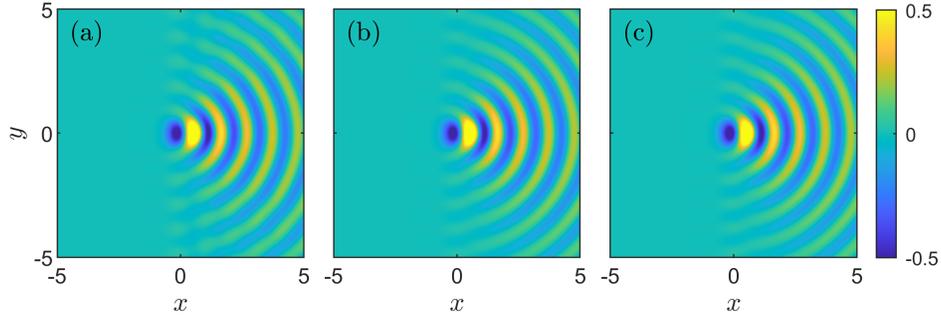

\centering
\psfragfig!{Picture/Final/WF_Comparison_solutions}
\caption{\label{Dipole waveforcing result} Pressure scaled by the maximum amplitude for the dipole test case computed using: (a) exact OWNS; (b) OWNS-P; (c) OWNS-R. Both OWNS-P and OWNS-R properly capture the one-way response to the dipole forcing.}
\end{figure}
\begin{figure}[!t]
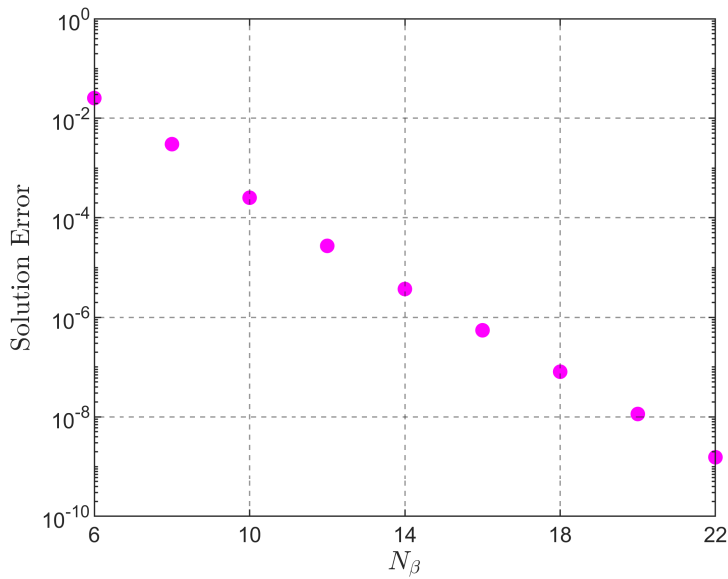

\centering
\psfragfig!{Picture/Final/WF_errorplot}
\caption{\label{Dipole waveforcing error}Solution error defined in (\ref{equation: solution error}) for the dipole problem. The error decreases exponentially with increasing number of recursion parameters.}
\end{figure}

The pressure fields obtained using the exact, OWNS-P, and OWNS-R projection operators are shown in Fig.~\ref{Dipole waveforcing result} for $N_{\beta} = 10 $.   OWNS-O and PSE are not applicable for the dipole problem due to their inability to accommodate the necessary forcing term.  Physically, the dipole forcing and its emitted acoustic waves are symmetric in the positive and negative $x$ directions, but here we are spatially integrating only in the positive $x$ direction.  Accordingly, the OWNS solutions capture rightgoing acoustic waves, as those propagating in the leftgoing direction are eliminated.  Visually, the three solutions are indistinguishable.  A more quantitative assessment of the error between the OWNS-R and exact OWNS solutions can be made using the solution error defined as 
\begin{equation} \label{equation: solution error}
     \frac{1}{A}\int_{A} \left| \frac{p_{OWNS}-p_{exact}}{p_{max}}\right|^{2} dA,
\end{equation}
where $p_{OWNS}$ is the pressure computed by OWNS-R, $p_{exact}$ is the pressure computed by the exact OWNS formulation, and $p_{max}$ is the maximum value of the pressure computed using the exact OWNS. The integration region $A$ is the area of the physical domain, not including the far-field damping layers. Fig.~\ref{Dipole waveforcing error} shows that the solution error decreases exponentially with increasing $N_{\beta}$ \cite{Towne2015one} and that reasonable accuracy can be achieved using $O(10)$ recursion parameters.  


\subsection{Turbulent jet}
 
Second, we consider the example of a turbulent jet.  Turbulent jets contain a diverse set of physical phenomena described by multiple modes of the linearized equations, making this a challenging test case and a useful point of comparison between OWNS and PSE.  Important types of waves supported by the linearized equations include the Kelvin-Helmholtz instability, which gives rise to large-scale coherent wavepacket structures \cite{Jordan2013wave}, the Orr mechanism \cite{Tissot2017sensitivity, Schmidt2018spectral}, and acoustic waves both trapped within the core of the jet \cite{Tam1989three, Towne2017acoustic} and emitted to the far field.  
 
We consider the specific case of a round jet with Mach number $M = U_{j} / c_{\infty} = 0.9$, Reynolds number $Re = \rho_{j} U_{j} D / \mu \approx 1 \times 10^{6}$, and temperature ratio $T_{j}/T_{\infty} = 1$, where the subscripts $j$ and $\infty$ denote conditions at the jet nozzle exit and in the far-field, respectively, and $D$ is the nozzle geometry.  The mean flow about which the Navier-Stokes equations are linearized is obtained from a high-fidelity large-eddy simulation \cite{Bres2018importance}.  Following Schmidt et al. \cite{Schmidt2018spectral}, we use a turbulent Reynolds number of $Re_T = 30000$ within the linearized equations.  This choice is motivated by recent work showing that using an eddy-viscosity model or reduced effective Reynolds number within linear models improves both near-field \cite{Pickering2021optimal} and far-field \cite{Pickering2021resolvent} predictions in free-shear flows.

The linearized equations in cylindrical coordinates are discretized in the radial direction using 200 grid points within a physical domain extending to $r/D = 10$ and an additional 80 grid points in the damping layer.  Since the jet is round, the mean flow is homogeneous in the azimuthal direction and the linearized equations can be decomposed into a series of independent azimuthal Fourier modes.  We focus on the asymmetric mode, which is typically of foremost interest in the study of jet aeroacoustics \cite{Cavalieri2012axisymmetric, Chen2021azimuthal}.  For the sake of brevity, we also focus on a single frequency coorsponding to a Stouhal number of $St = \omega D / 2\pi U_{j} = 0.35$, which is close to the most unstable frequency for the Kelvin-Helmholtz mode and the same condition used by Towne et al. \cite{Towne2015one} to demonstrate the OWNS-O method.  The equations are integrated from $x/D=0.5$ to $x/D=25$.  All OWNS solutions use a second-order diagonally implicit Runge-Kutta method \cite{Hairer1996solving} with a uniform step size of $ \Delta x = 0.1$. The step size for PSE is set to its minimum stable value \cite{Li1996on, towne2019critical} (plus a small safety factor) and is on average about four times larger than the OWNS step size.  We set $N_\beta = 20$ for all OWNS calculations.  To enable comparisons with PSE and OWNS-O, we excite the flow using an initial perturbation at the near-nozzle boundary, corresponding to the local Kelvin-Helmholtz eigenmode, and do not consider a volumetric forcing.

\begin{figure}[!t]
\centering
\psfragfig!{Picture/Final/TJ_Comparison_Fc1}
\caption{\label{linear wavepacket correction2, filters, contour plot} Eigen-space results for the jet test case for: (a) $N_{\beta}$ = 10; (b) $N_{\beta}$ = 15; (c) $N_{\beta}$ = 20; (d) $N_{\beta}$ = 25.  Symbols: (\lowersym{\textcolor{magenta}{\bf\SmallCross}}) eigenvalues of $\boldsymbol{M}$; (\lowersym{\textcolor{blue}{\bf\SmallCircle}}) rightgoing eigenvalues of $\boldsymbol{P}_{N_\beta}\boldsymbol{M}$; (\lowersym{\textcolor{red}{\bf\SmallCircle}}) leftgoing eigenvalues of $\boldsymbol{P}_{N_\beta}\boldsymbol{M}$; (\lowersym{\textcolor{green}{\bf\FilledSmallSquare}}) ${\beta_j}^{+}$; (\lowersym{\textcolor{black}{\bf\FilledSmallSquare}}) ${\beta_j}^{-}$. The contours show the magnitude of $\mathcal{E}(\alpha)$, which converges toward 1 and 0 in the vicinity of rightgoing and leftgoing modes, respectively.}
\end{figure}

Fig.~\ref{linear wavepacket correction2, filters, contour plot} shows eigen-space results at $x/D=1$ for the jet analogous to those shown in Fig.~\ref{Waveforcing, correction2, PM} for the dipole problem.  The pink (x) symbols show a subset of the eigenvalues of $\boldsymbol{M}$ for the jet.  The L-shaped branches that extend along the imaginary axis toward plus and minus infinity are rightgoing and leftgoing free-stream acoustic modes analogous to those observed in the dipole problem.  The discrete mode located at approximately $(3,-1)$ in the complex plane is the Kelvin-Helmholtz mode.  The branch beginning at around $(2,0)$ and extending upward into the positive imaginary plane superimpose non-normally to generate Orr waves \cite{Tissot2017sensitivity}.  Finally, the discrete modes at about $(-9, \pm4)$ are acoustic waves trapped within the core of the jet, which are unimportant at this frequency \cite{Towne2017acoustic}.  The green and back squares indicate the locations of the $\beta^{+}_{j}$ and $\beta^{-}_{j}$ recursion parameters, respectively.  The background contours show the magnitude of $\mathcal{E}(\alpha)$ for $N_{\beta} = 10, 15, 20$ and $25$ in the four subfigures, and it is again clear that this function approaches zero and one for regions of the complex plane occupied by leftgoing and rightgoing modes, respectively, as desired.  Finally, the blue and red circles show that the rightgoing and leftgoing eigenvalues of the one-way operator are converging toward the eigenvalues of $\boldsymbol{M}$ and toward zero, respectively, as the order of the recursions is increased, with good accuracy observed for the two higher values.

\begin{figure}[!t]
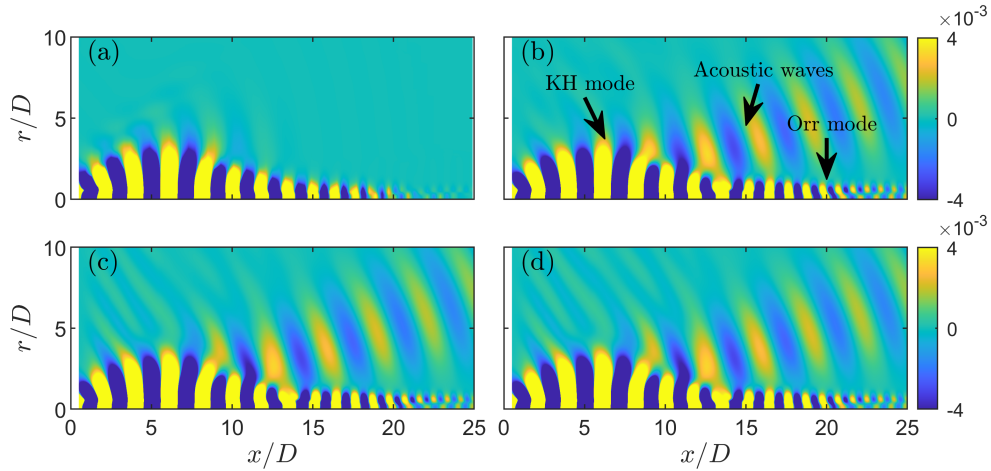

\centering
\psfragfig!{Picture/Final/TJ_Comparison_solutions}
\caption{\label{linear wavepacket correction2 and other methods.}Contours of pressure fluctuations for the jet scaled by the maximum amplitude: (a) PSE; (b) OWNS-O; (c) OWNS-P; (d) OWNS-R. PSE fails to capture the Orr waves and radiated acoustic waves captured by all three OWNS variants.}
\end{figure}

Fig.~\ref{linear wavepacket correction2 and other methods.} shows the pressure fields calculated by PSE and the OWNS-O, OWNS-P, and OWNS-R methods for $N_\beta = 20$. All four solutions contain a clear wavepacket structure, produced by the Kelvin-Helmholtz mode, in the jet near field that begins at the computational inlet and persists to $x/D \approx 20$.  This is the dominant mode tracked by PSE.  However, the PSE solution misses two other important features that are properly captures by all three OWNS solutions.  The first is the acoustic waves that are emitted from the Kelvin-Helmholtz wavepacket.  
These acoustic waves are described by rightgoing modes whose eigenvalues lie in a different region of the complex plane, as shown in Fig.~\ref{linear wavepacket correction2, filters, contour plot}, and therefore cannot be captured by PSE \cite{towne2019critical}.  However, these modes are faithfully retained by all variants of OWNS, allowing these methods to properly capture the acoustic radiation associated with the Kelvin-Helmholtz instability, which is a primary source of jet noise \cite{Jordan2013wave}.  The second feature captured by OWNS but not PSE is Orr waves.  These waves are less visually obvious and can be easily mistaken as an extension of the Kelvin-Helmholtz wavepackets.  However, the higher amplitude of the waves in the OWNS solutions in the region $15<x/D<20$ and their persistence beyond $x/D=20$ is a result not of the Kelvin-Helmholtz mode, but of a superposition of multiple non-normal stable modes that together make up the tilted structures typical of the Orr mechanism \cite{Tissot2017sensitivity}.  Again, PSE cannot capture these Orr waves due to their reliance on multiple eigenvalues that differ from the Kelvin-Helmholtz mode that PSE is tracking.  

A more quantitative assessment of these observations is provided in Fig~\ref{linear wavepackets, spatial marching result for correction2-line result}, where the pressure from the PSE and three OWNS solutions is plotted as a function of $x/D$ at two radial positions, $r/D = 0.5$ and 5.  Consider first the results for $r/D = 0.5$ (the lipline of the jet) shown in Fig.~\ref{linear wavepackets, spatial marching result for correction2-line result}(a).  While the three OWNS solutions are indistinguishable, the PSE solution differs in two ways.  Small but discernible differences can be observed between the Kelvin-Helmholtz wavepackets ($0<x/D<15$) computed by PSE and OWNS.  More importantly, the OWNS solutions capture low-amplitude Orr waves in the region $x/D>15$ that PSE is unable to capture.  The amplitude of these Orr waves is low in this test case because we have excited to system using only an initial condition rather than a volumetric forcing (which was necessary to enable PSE and OWNS-O solutions), while Orr waves are most efficiently excited by volumetric forcing from turbulence \cite{Schmidt2018spectral}.  Including such forcing is critical for creating accurate jet noise models \cite{Towne2017statistical, Towne2018spectral}, making the ability of OWNS to capture these waves vital for this endeavor \cite{Towne2014continued, Rigas2017oneway}.

\begin{figure}[!t]
    \centering
    \begin{overpic}{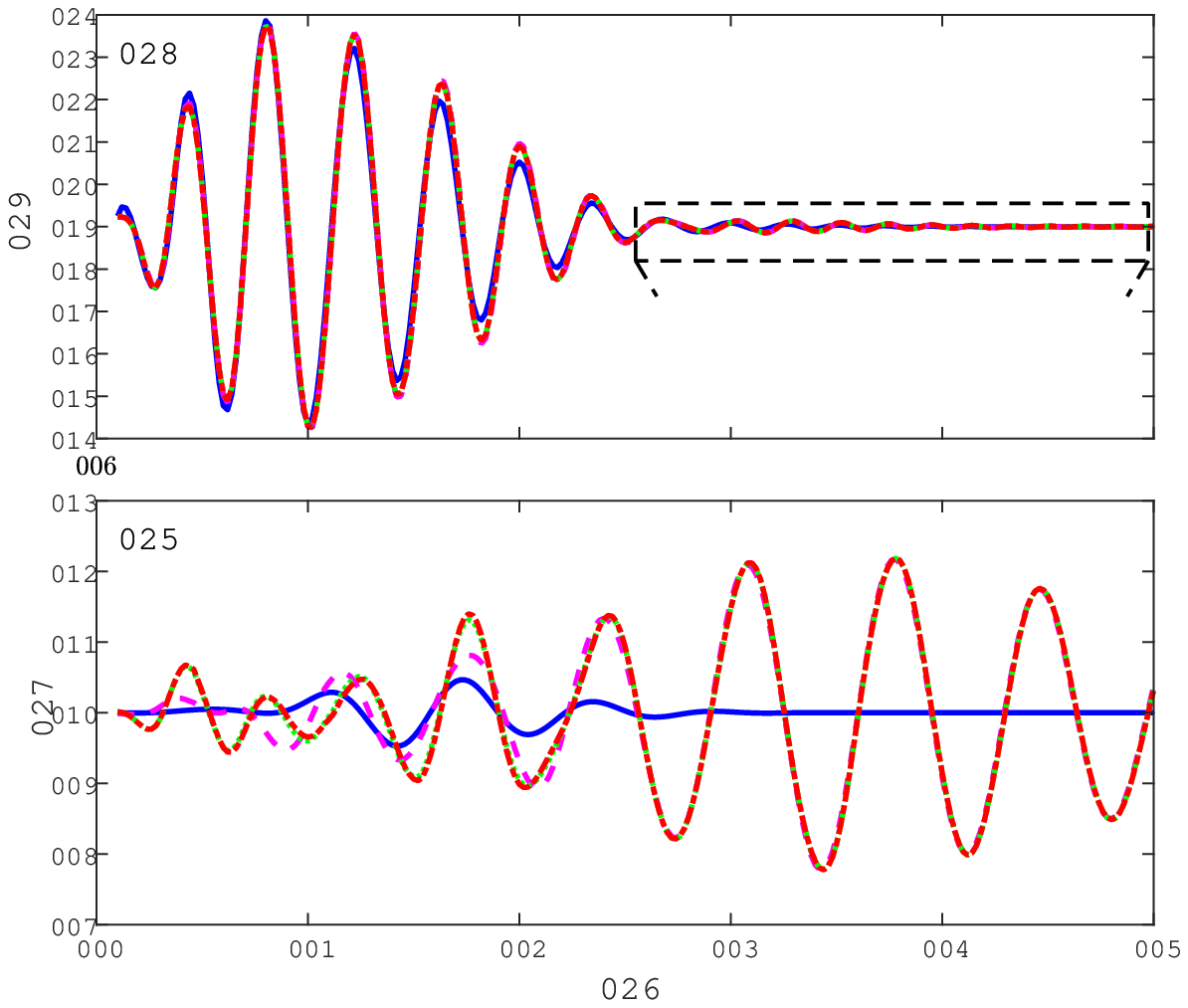}
    \put(53,44.5){\frame{\includegraphics[scale = 0.61]{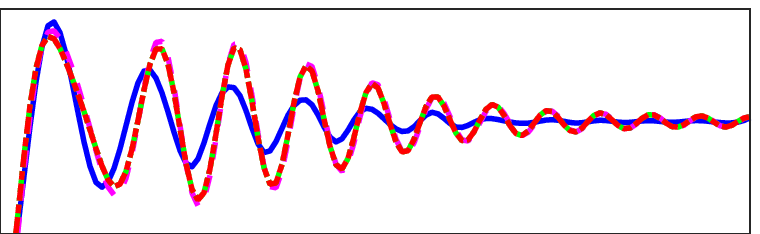}}}
    \end{overpic}
    \caption{Pressure fluctuations, scaled by the maximum magnitude, in the jet at (a) $r/D=0.5$ and (b) $r/D=5$ computed using: (\textcolor{blue}{—}) PSE;  (\textcolor{magenta}{- -}) OWNS-O; (\textcolor{green}{- -}) OWNS-P; (\textcolor{red}{- -}) OWNS-R.  All three OWNS variants give nearly the same pressure fields, while the result of PSE differ due to missing Orr waves in (a) and radiated acoustic waves in (b).} 
    \label{linear wavepackets, spatial marching result for correction2-line result}
\end{figure}

Fig.~\ref{linear wavepackets, spatial marching result for correction2-line result}(b) shows the pressure for the four methods along $r/D = 5$.  For $x/D \lesssim 15$, the pressure field at this radial distance contains both hydrodynamic and acoustic components \cite{Suzuki2006instability}; the presence of remnant hydrodynamic contributions (whose amplitude decays exponentially with increasing $r/D$) is made clear by the non-zero PSE solution in this region.  The three OWNS solutions contain additional acoustic contributions, leading to a higher amplitude than PSE.  The OWNS-P and OWNS-R solutions are closely matched, as expected, but differ somewhat from the OWNS-O solution in this region.  A close comparison between the three OWNS solutions in Fig.~\ref{linear wavepacket correction2 and other methods.} reveals that the higher amplitude of the OWNS-P and OWNS-R solutions is the result of additional high-angle, low amplitude acoustic waves that are present in these solutions but not in the OWNS-O solution.  This minor discrepancy was previous observed in comparisons between OWNS-O and OWNS-P and the root cause is under further investigation, but we emphasize that these extra acoustic waves are weak and dissipate quickly compared to the main acoustic beam that radiates at lower angles to the jet axis.  For $x/D 	\gtrsim 15$, the solution at $r/D=5$ is dominated by this main acoustic beam.  All three OWNS solutions capture this low-angle acoustic radiation, which is the target of interest in jet aeroacoustics studies, while the PSE solution completely misses this radiation due to its inability to capture acoustic modes during the spatial march \cite{towne2019critical}.  

Table~\ref{Turbulent jet, time cost for correction2} shows the computational cost of the different methods.  First comparing between the three OWNS methods, we see that OWNS-R is approximately 7 and 29 times faster than OWNS-O and OWNS-P, respectively.  The most relevant comparison is between OWNS-R and OWNS-P, since these two methods share the same capabilities, i.e., the ability to accommodate volumetric forcing. 
To aid in making cost comparisons between OWNS and PSE, we report both the wall time of the simulations as well as the wall time per step in the spatial march.  These two metrics convey different information because the large step size needed to stablize the PSE march leads to a smaller number of total steps.  While taking a smaller number of steps may at first glance seem to be a feature of PSE, in reality it is a undesirable artifact of the ad-hoc PSE parabolization.  In other words, if we could take smaller steps in the PSE march, we would, and indeed several methods have been proposed to enable smaller steps by reducing the minimum step size restriction  \cite{Chang1991compressible, HajHariri1994characterisitics, Li1996on, Andersson1998on}.  Stated differently, the OWNS solutions could also be computed using larger, PSE-like step size, but this leads to a deleterious loss of accuracy, just as it does for PSE.  Given that the smaller number of steps is actually a detrimental limitation of PSE, cost \emph{per step} is arguably the more relevant metric for comparing PSE and OWNS.  Using this metric, the OWNS-O and OWNS-P solutions are approximately 21 and 92 times more expensive than PSE.  In contrast, OWNS-R is only 3 times more expensive, while delivering a solution equivalent to OWNS-P.  Thus, our new OWNS formulation approaches the cost of PSE while retaining the accuracy of previous OWNS variants. The cost of all three OWNS solutions could be further reduced by using a linear multistep method rather than the two-stage Runge-Kutta method used here. With respect to memory cost, OWNS-O and OWNS-P use approximately 29 and 85 times more RAM than PSE. This is consistent with the fact that OWNS-O and OWNS-P need to form and solve a linear system with the size of $NN_\beta$. In contrast, the  memory cost for OWNS-R is the same as PSE. 

\begin{figure}[!t]
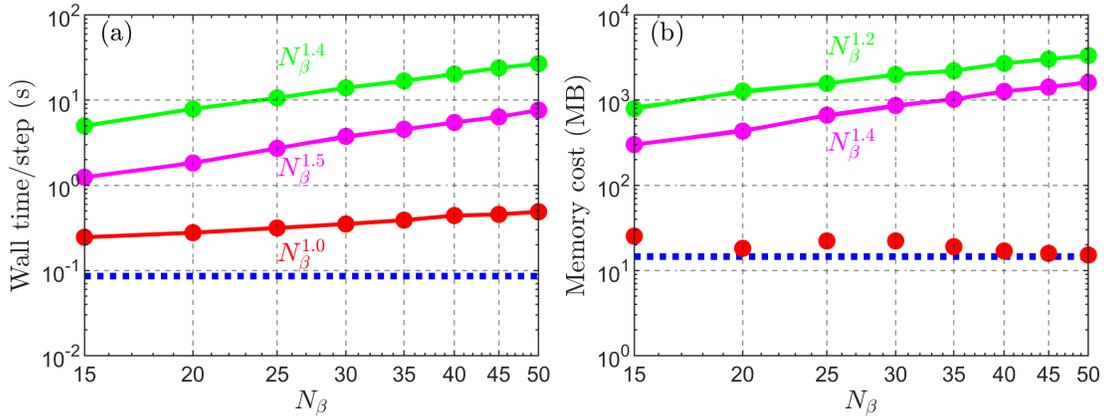

\centering
\psfragfig!{Picture/Final/TJ_time_cost}
\caption{\label{TJ cost}Wall time / step and memory cost as a function of the recursion order $N_{\beta}$:  (\tiny \lowersym{\textcolor{blue}{\FilledSmallSquare}}\lowersym{\textcolor{blue}{\FilledSmallSquare}}\lowersym{\textcolor{blue}{\FilledSmallSquare}}\footnotesize) PSE; (\lowersym{\textcolor{magenta}{\bf\FilledSmallCircle}}) OWNS-O; (\lowersym{\textcolor{green}{\bf\FilledSmallCircle}}) OWNS-P;  (\lowersym{\textcolor{red}{\bf\FilledSmallCircle}}) OWNS-R; (\textcolor{magenta}{—}) fit line for OWNS-O; (\textcolor{green}{—}) fit line for OWNS-P; (\textcolor{red}{—})fit line for OWNS-R.  The scalings of fit lines are labelled and the results match the theoretical prediction reported in Table~\ref{tab:theoretical_scaling}.  Note that the cost of PSE is independent of $N_{\beta}$ since it does not involve recursions, and the memory cost for OWNS-R is nearly independent of $N_\beta$, so the fit line is not shown here.}
\end{figure}

Finally, the computational costs for the PSE, OWNS-O, OWNS-P, and OWNS-R solutions, measured in terms of wall-time/step and maximum memory usage, are reported in Fig.~\ref{TJ cost} as a function of $N_{\beta}$.  The cost of PSE is not a function of $N_{\beta}$ since the PSE procedure does not involve recursion parameters. The OWNS-O and OWNS-P methods exhibits wall time scalings of $O(N_{\beta}^{a})$ with $a\approx 1.5$ and $1.4$, respectively, while OWNS-R achieves the expected linear scaling, $a\approx 1.0$.  The memory scaling of the OWNS-O and OWNS-P methods is $O(N_{\beta}^{b})$ with $b\approx 1.4$ and $1.2$, respectively, whereas the memory consumption of OWNS-R is independent of $N_{\beta}$ by construction.  Moreover, the maximum memory consumption of OWNS-R is nearly identical to that of PSE.  Overall, these results confirm the theoretical cost scaling analysis reported in Table \ref{tab:theoretical_scaling} and highlight the significant computational advantages of OWNS-R.


\setlength{\tabcolsep}{24pt}
\begin{table}
\small
\centering
\caption{\footnotesize Computational cost for PSE, OWNS-O, OWNS-P and OWNS-R for the turbulent jet case.}
\label{Turbulent jet, time cost for correction2}
\renewcommand\arraystretch{1.33}
\setlength\tabcolsep{10pt}
\begin{tabular}{lllll}
\hline
\textbf{Method} & \textbf{Wall time (s)}& \textbf{Steps} ($N_{x})$ &\textbf{Wall time / step (s)}  & \textbf{Memory cost (MB)}   \\ 
\hline

PSE             & 5.4      & 63        &  0.086  & 15\\ 

OWNS-O          & 449       & 246       & 1.83     & 432\\ 

OWNS-P          & 1941      & 246       & 7.9     & 1278\\ 

OWNS-R          & 68      & 246       & 0.28  & 18   \\
\hline
\end{tabular}
\end{table}

\subsection{Supersonic boundary layer}

Third, we consider a supersonic boundary layer to demonstrate OWNS-R applied to a wall-bounded flow.  Modeling the amplification of disturbances in supersonic boundary layers is important for predicting laminar to turbulent transition, which is critical for managing aerodynamic and thermal loads in high-speed flight.  

Following Ma \& Zhong \cite{ma2003receptivity}, we consider a Mach 4.5 boundary layer over an adiabatic flat plate. The flow conditions are: $M_{\infty} = 4.5$, $T_{\infty}^* = 65.15K$, $p_{\infty}^*=728.44Pa$, $Pr = 0.72$, unit Reynolds number $Re_{\infty}^*=\frac{\rho_{\infty}^*u_{\infty}^*}{\mu_{\infty}^*}=7.2\times10^6$ $m^{-1}$. The base flow about which the Navier-Stokes equations (written in Cartesian coordinates) are linearized is obtained from the Howarth–Dorodnitsyn transformation of the compressible Blasius boundary layer equations under the assumption of zero pressure gradient.  We focus on spanwise-constant perturbations such that the linearized equations reduce to their two-dimensional form. 

 In the streamwise direction, the various OWNS equations are integrated from $R = 400$ to $1200$, where $R = \sqrt{Re_x}$ and $ Re_x = Re_{\infty}^*x^*$. Here, $R$ is the local Reynolds number, $Re_x$ is the dimensionless local Reynolds number, and $x^*$ is the dimensional coordinate in meters measured from the leading edge along the plate surface \cite{ma2003receptivity}. The wall-normal coordinate $y$ is discretized using 200 points within the physical domain extending to $Y = 1200$. Here, $Y$ is the dimensionless wall-normal direction, non-dimensionalized using the boundary thickness at $R = 400$.   All results shown below are for the non-dimensional frequency $F = \omega^* \mu_{\infty}^* /(\rho_{\infty}^* u_{\infty}^{*2}) = 2.2 \times 10^{-4}$, following one of the cases in Ma \& Zhong \cite{ma2003receptivity}.

\begin{figure}[!t]
\centering
\begin{overpic}{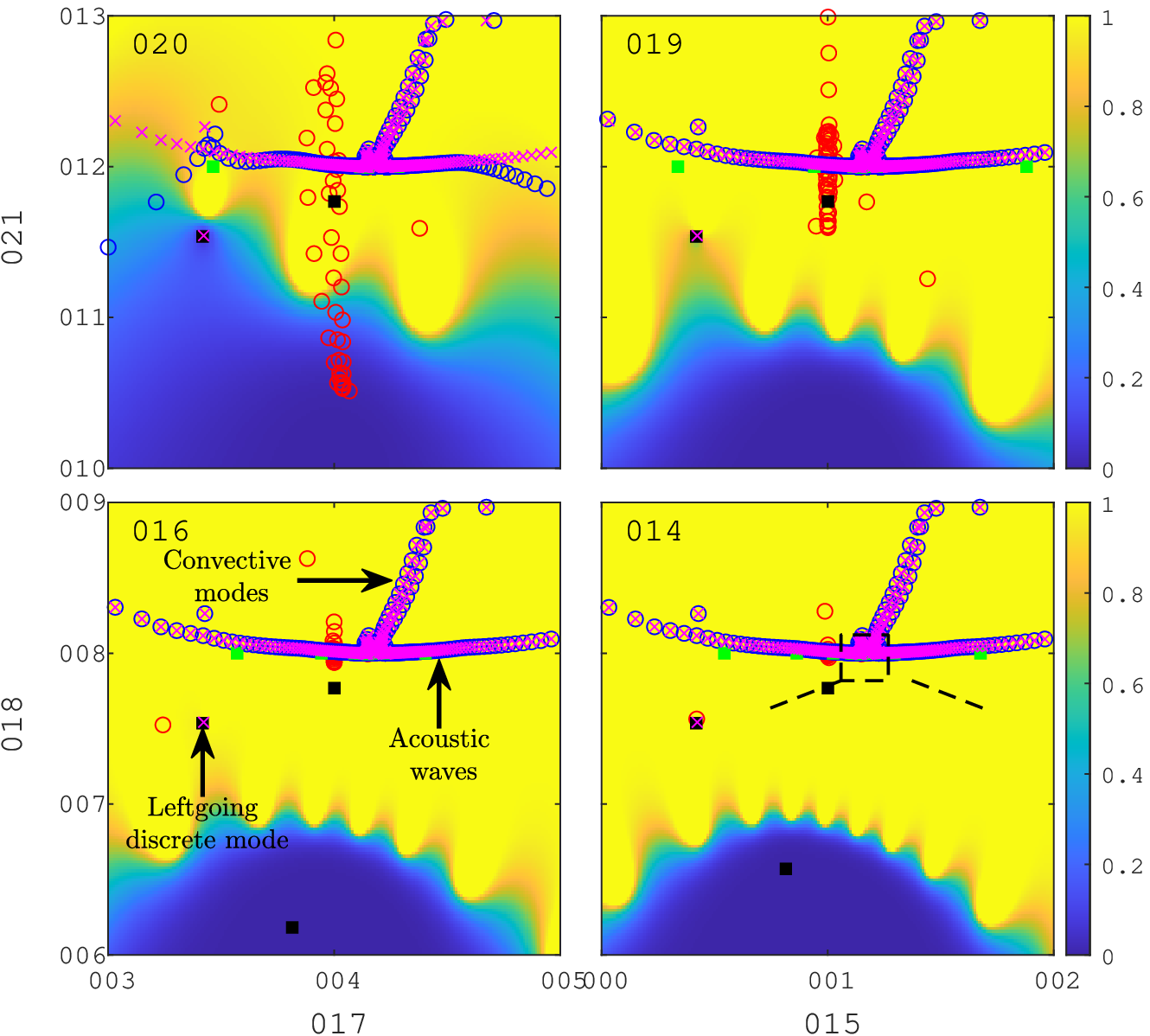}
    \put(68,9.7){\frame{\includegraphics[scale = 0.185]{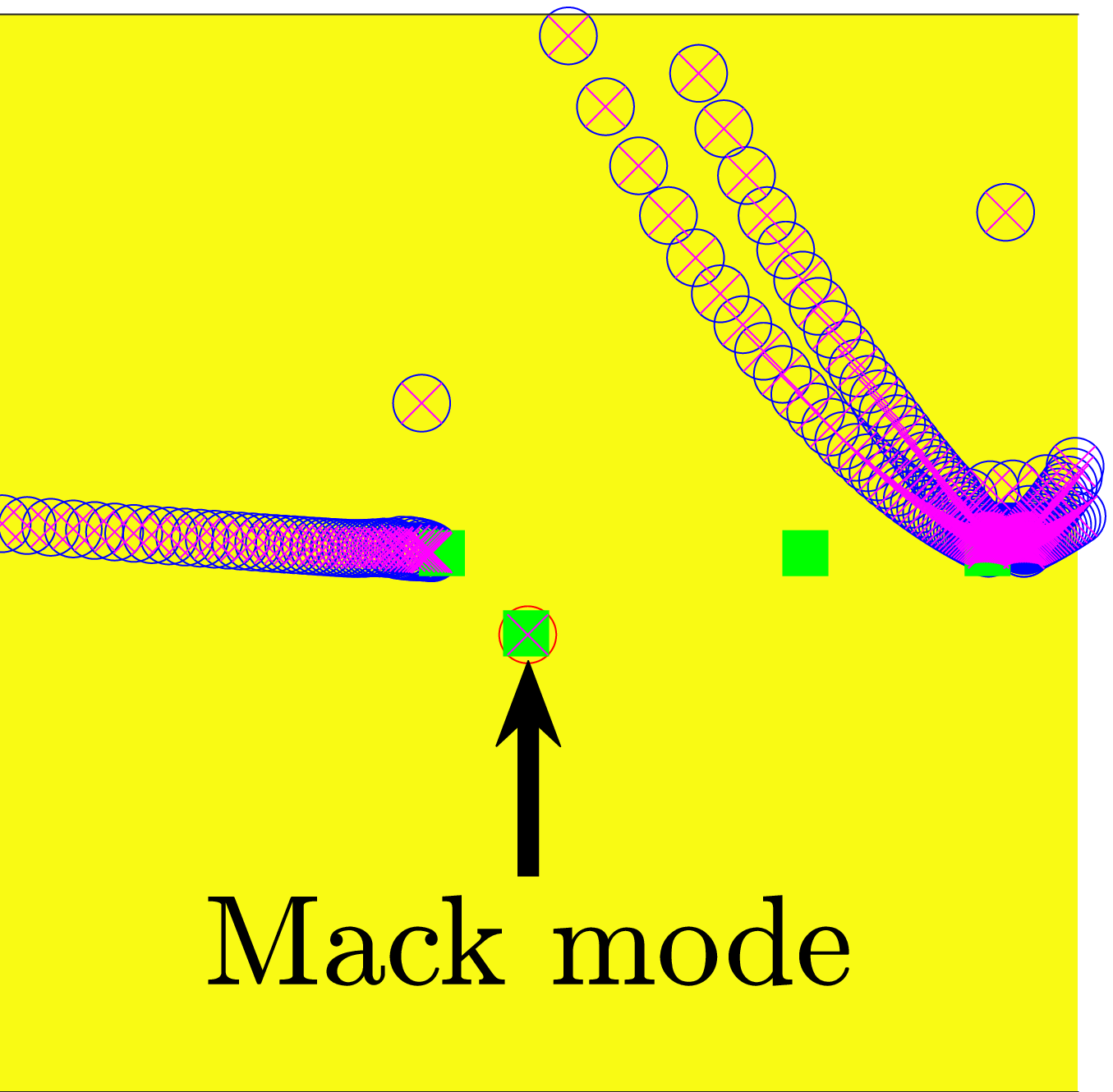}}}
    \end{overpic}
\caption{\label{fig:BL_eigs} Eigen-space results for the supersonic boundary layer test case for: (a) $N_{\beta}$ = 10; (b) $N_{\beta}$ = 15; (c) $N_{\beta}$ = 20; (d) $N_{\beta}$ = 25.  Symbols: (\lowersym{\textcolor{magenta}{\bf\SmallCross}}) eigenvalues of $\boldsymbol{M}$; (\lowersym{\textcolor{blue}{\bf\SmallCircle}}) rightgoing eigenvalues of $\boldsymbol{P}_{N_\beta}\boldsymbol{M}$; (\lowersym{\textcolor{red}{\bf\SmallCircle}}) leftgoing eigenvalues of $\boldsymbol{P}_{N_\beta}\boldsymbol{M}$; (\lowersym{\textcolor{green}{\bf\FilledSmallSquare}}) ${\beta_j}^{+}$; (\lowersym{\textcolor{black}{\bf\FilledSmallSquare}}) ${\beta_j}^{-}$. The contours show the magnitude of $\mathcal{E}(\alpha)$.}
\end{figure}

Fig.~\ref{fig:BL_eigs} shows the same visualization of the OWNS-R projection at $R = 400$ as discussed for the previous two examples.  Several different types of eigenvalues of $\boldsymbol{M}$ can be identified.  First, the horizontal branch of modes represents free-stream acoustic waves \cite{Li1996on}; this branch looks different from the free-stream acoustic branches in the jet problem because of the supersonic free-stream velocity above the boundary layer.  Second, the vertical branch that begins near the origin and extends into the positive complex plane represents stable, convective vorticity and entropy modes. Third, the unstable discrete mode is a Mack mode, which is an important instability mode in supersonic boundary layers \cite{Knisely2019sound, Fedorov2011transition, Fedorov2011high}. All of these modes discussed so far are rightgoing modes, and thus should be preserved in the OWNS march.  Finally, one leftgoing discrete mode is visible \cite{Li1996on}, and several more exist outside the axes of the figure; these modes must be eliminated to achieve a stable march.  As before, the black and green squares show locations of the $\beta_{j}^{+}$ and $\beta_{j}^{-}$ recursion parameters, respectively, and the background contours show the value of the function $\mathcal{E}(\alpha)$ for $N_{\beta} = 10, 15, 20$, and $25$ in the four subfigures.  Clearly, $\mathcal{E}(\alpha)$ is achieving its desired limits with increased accuracy as $N_{\beta}$ is increased, especially for the unstable discrete mode.

The unstable Mach mode is used to initialize the PSE and OWNS marches at $R = 400$.  Fig.~\ref{supersonic BL, spatial marching result for correction2} shows the real part of the pressure field obtained using PSE and OWNS-R with $N_{\beta} = 20$; the OWNS-O and OWNS-P solutions are indistinguishable from the OWNS-R solution and are thus omitted for brevity.  While both the PSE and OWNS-R solutions contain a qualitatively similar structure close to the wall, the latter includes a more complex pattern further from the wall due to additional acoustic waves that cannot be captured by PSE.  

\begin{figure}[!t]
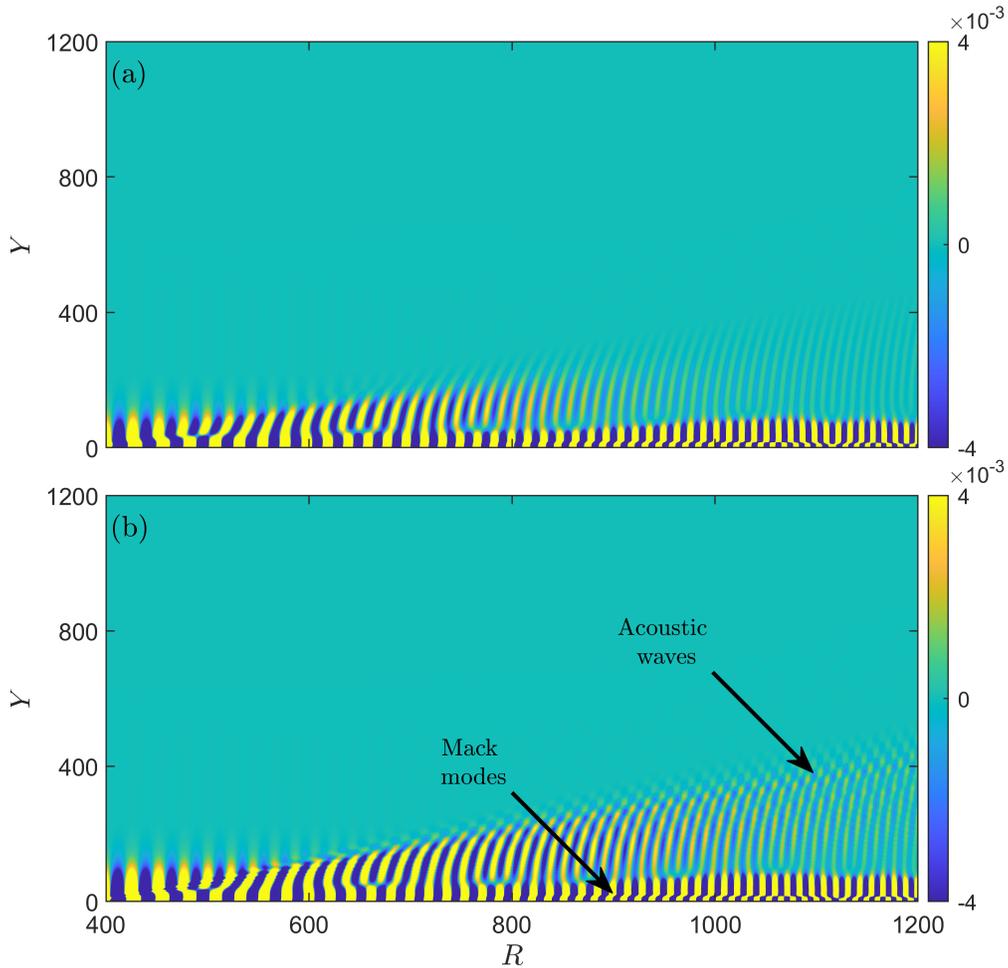

    \centering
    \psfragfig!{Picture/Final/HB_Comparison_solutions2}
    \caption{Contours of pressure fluctuations in the supersonic boundary layer scaled by the maximum amplitude: (a) PSE; (b) OWNS-R. Both methods capture the Mack mode, but PSE fails to capture the radiated acoustic waves.} 
    \label{supersonic BL, spatial marching result for correction2}
\end{figure}

Fig.~\ref{supersonic BL, spatial marching result for correction2-line result} more clearly shows the discrepancies in the PSE solution by plotting the pressure amplitude as a function of $x^{*}$ for two wall-normal positions, $Y=0$ and $300$. For reference, the pressure along the wall from the DNS results of Ma \& Zhong \cite{ma2003receptivity} is also shown. Note that the solutions for all three OWNS variants lie on top of one another. Along the wall, shown in Fig. \ref{supersonic BL, spatial marching result for correction2-line result}(a), the peak of the pressure fluctuations is slightly further downstream in the DNS results ($x^{*} = 0.155m$) compared to PSE ($x^* = 0.147m$) and OWNS ($x^* = 0.145m$). This can be attributed to differences in the DNS mean velocity compared to the similarity transformation used for our calculations.  Specifically, the momentum thickness growth as the boundary layer develops downstream if slightly different in the DNS compared to the similarity solution.  These differences aside, it is still clear that the OWNS solutions do a better job than PSE in capturing the growth at intermediate $x^{*}$ values.  The benefit of OWNS over PSE is more obvious at the position further from the wall, shown in Fig.~\ref{supersonic BL, spatial marching result for correction2-line result}(b).  The OWNS solutions contain a clear amplitude modulation caused by the acoustic waves that are emitted from the main instability wave; PSE misses this due to its inability to properly capture the additional waves involved in this physical phenomena.

\begin{figure}[!t]
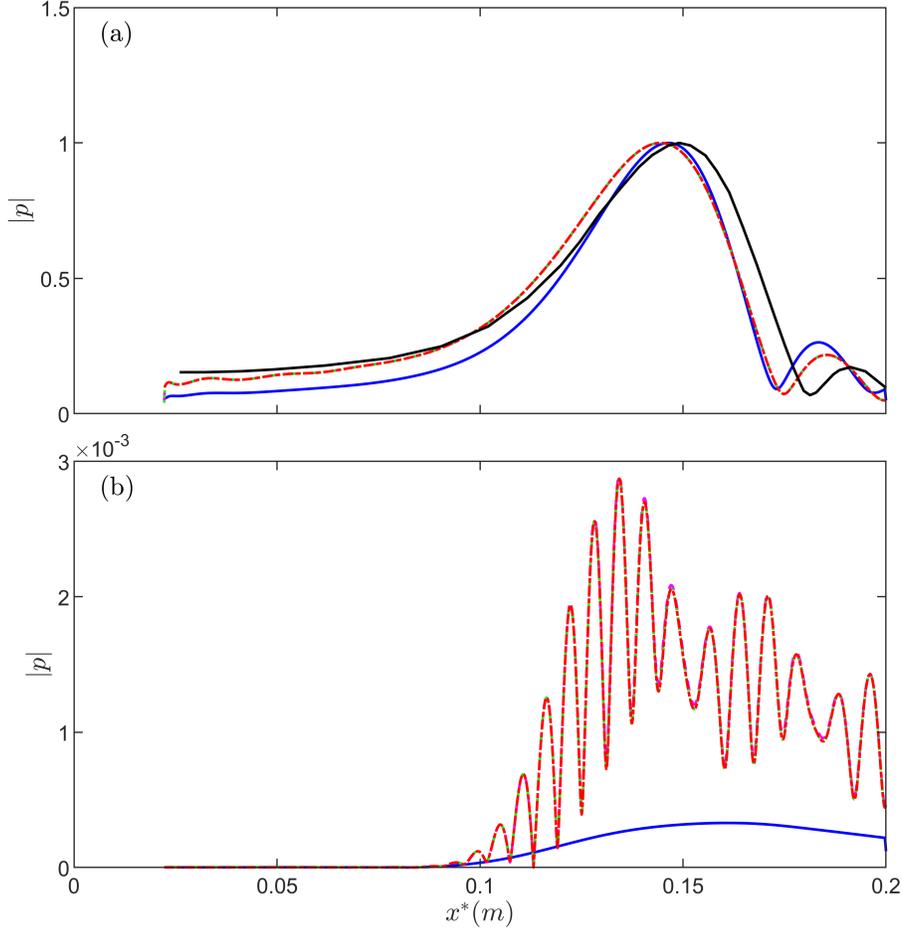

    \centering
    \psfragfig!{Picture/Final/HB_Comparison_solutions4}
    \caption{Pressure amplitude in the supersonic boundary layer (scaled by the maximum) at (a) $Y = 0$ and (b) $Y = 300$, computed using: (\textcolor{blue}{—}) PSE;  (\textcolor{magenta}{- -}) OWNS-O; (\textcolor{green}{- -}) OWNS-P; (\textcolor{red}{- -}) OWNS-R; (\textcolor{black}{—}) DNS from Ma \& Zhong \cite{ma2003receptivity}.  All OWNS variants capture the radiated acoustic waves, unlike PSE. } 
    \label{supersonic BL, spatial marching result for correction2-line result}
\end{figure}

Finally, Table \ref{supersonic BL, time cost for correction2} shows that OWNS-R obtains these superior results at a CPU cost per step of only two times that of PSE and using equal (or actually sightly lower) memory.  Moreover, OWNS-R obtains a solution indiscernible from the OWNS-O and OWNS-P solutions, but 10 and 20 times faster using 15 and 32 times less memory, respectively.

\setlength{\tabcolsep}{24pt}
\begin{table}[t]
\small
\centering
\caption{\footnotesize Computational cost for PSE, OWNS-O, OWNS-P and OWNS-R for supersonic boundary layer case.}
\label{supersonic BL, time cost for correction2}
\renewcommand\arraystretch{1.33}
\setlength\tabcolsep{10pt}
\begin{tabular}{lllll}
\hline
\textbf{Method} & \textbf{Wall time (s)}                & \textbf{Steps} ($N_{x})$         &   \textbf{Wall time / step (s)} & \textbf{Memory cost (MB)}     \\ 
\hline

PSE             & 17.7     & 323       & 0.055 & 46  \\ 
OWNS-O          & 1099     & 1201       & 0.92 & 458    \\ 

OWNS-P          & 2541      & 1201       & 2.12 & 995 \\ 

OWNS-R          & 154     & 1201       & 0.128 & 31   \\
\hline
\end{tabular}
\end{table}


\section{Conclusions} 
\label{sec:conclusions}
 
In this paper, we have developed a new variant of the one-way Navier-Stokes equations, which we call OWNS-R.  As in previous versions of OWNS, the basic idea is to evolve rightgoing modes, i.e., those that transfer energy in the downstream direction, via spatial integration of the governing equations in the downstream direction.  The OWNS-R method is formulated in terms of a projection operator $\boldsymbol{P}$ that eliminates leftgoing modes, i.e., those that transfer energy in the upstream direction, without altering rightgoing modes.  As the flow equations are integrated in the downstream direction, this projection operator is applied to the state vector after each step in the spatial march to remove the leftgoing modes and prevent them from destabilizing the march.  The method is derived in terms of the eigen-decomposition of the spatial LNS operator $\boldsymbol{M}$, but actually implementing it in this manner is unnecessarily expensive.  Instead, the action of the projection operator on a vector is efficiently approximated using a set of recursion equations similar to those employed by previous OWNS variants.  Unlike these previous methods, the OWNS-R recursion equations can actually be solved recursively, i.e., one at a time, rather than as a coupled set.  This leads to improved cost scaling with respect to the order of the recursions $N_{\beta}$: the CPU cost of OWNS-R scales linearly with $N_{\beta}$ and its memory usage in independent of $N_{\beta}$.  We show by comparing these scaling results with other methods and via example problems that OWNS-R is an order of magnitude less costly than previous OWNS variants and similar in cost to PSE.  However, while PSE can capture only one rightgoing mode, OWNS-R properly evolves all rightgoing modes supported by the spatial LNS operator.  

Since both OWNS-R and OWNS-P (a previous variant of OWNS) are formulated in terms of projection operators, we conduct a detailed comparison between these two methods.  Both methods approximate the same exact projection operator and converge to this common limit as the order of the recursions increases.  Compared to the exact projection operator, the eigenvalues of the OWNS-P projection operator are exact but its eigenvectors are approximated.  Conversely, the eigenvalues of the OWNS-R projection operator are approximated while its eigenvectors are exact.  A notable benefit of exact eigenvectors is that expressions for the OWNS-R projection error can be derived that enable \emph{a priori} error analysis as a function of the position in the complex plane of an eigenvalue of interest and a chosen set of recursions parameters without any knowledge of the corresponding eigenvector.  Additionally, whereas poor convergence of the function $\mathcal{F}(\alpha)$ for one mode potentially impacts the projection error of all modes for OWNS-O and OWNS-P, the error in each mode is completely uncoupled from all other modes in OWNS-R.  

Critically, recursion parameters that have been derived for a variety of flows \cite{Towne2013improved, Towne2014continued, Towne2015one, Rigas2017stability, kamal2020application} for use with OWNS-O and OWNS-P can also be used for OWNS-R.  The error decoupling described above could also provide additional flexibility in defining the recursions parameters for OWNS-R.  For example, whereas recursion parameters must be defined such that exactly $N_{+}$ modes are retained in OWNS-O and OWNS-P, unimportant rightgoing modes could be projected out in OWNS-R, which could simplify the choice of recursions parameters.

We also use the projection operator differently than in OWNS-P.  There, the projection operator is used to project the spatial LNS equations, the projected equations are then integrated, and the resulting state vector is again projected after each step in the march to prevent buildup of energy in leftgoing modes.  In OWNS-R, we propose to skip the intermediate step of projecting the equations and instead rely exclusively on projecting the state vector after each step in the march.  We show that both approaches are nearly equivalent, ensuring stability of the OWNS-R march.

The capabilities of our OWNS-R method are demonstrated using three example problems.  First, we consider a simple case in which a dipole forcing term excites acoustic waves in a quiescent fluid.  This provides a clean environment to illustrate the properties of the method and demonstrate the convergence of the eigenvalues of the approximate projection operator and of the solution. Second, we consider a prototypical free-shear flow -- a Mach 0.9 turbulent jet.  Whereas PSE can only capture the Kelvin-Helmholtz instability of the annular shear layer, all three OWNS variants additional capture Orr waves and radiated acoustic waves. The improved cost scaling of OWNS-R compared to OWNS-O and OWNS-P is also confirmed.  Third, we consider a prototypical wall-bounded flow -- a Mach 4.5 flow over an adiabatic flat plate. In this case, the dominant instability mode is a Mack mode and both PSE and all three OWNS variants capture this mode. The OWNS methods also capture the radiated acoustic waves..

The low-cost, accurate solutions offered by OWNS-R make it an attractive candidate to replace global methods, previous OWNS variants, and PSE in a variety of contexts.  Traditional stability analyses of slowly varying flows can be conducted using OWNS-R at lower cost than OWNS-O or OWNS-P and with improved accuracy compared to PSE, specifically the ability to capture the influence of multiple modes.  Reducing the cost of resolvent analysis has recently received considerable attention \cite{Ribeiro2020randomized, Martini2021efficient, Farghadan2021randomized}, and OWNS-R could be used in place of OWNS-P to further reduce the CPU and memory cost of the approach recently developed by Towne et al. \cite{Towne2021fast} and Rigas et al. \cite{Rigas2021fast} for slowly varying flows.  Finally, since OWNS-R can accommodate a forcing term, it could be used as the basis for a nonlinear extension of OWNS analogous to nonlinear PSE.  Whereas nonlinear PSE retains nonlinear interactions between just one mode at each frequency, nonlinear OWNS could capture nonlinear interactions between all rightgoing modes at minimal additional cost.

 
 \section*{Acknowledgements}
 
 The authors thank Georgios Rigas, Imperial College London, for supplying the base flow code for the supersonic boundary layer test case.


\begin{appendices}
 
 \section{Extension to singular \texorpdfstring{$\boldsymbol{A}$}{\bf \textit{A}}}

 \label{App:singularA}
 
 We assumed in the main text that the matrix $\boldsymbol{A}$ is invertible, i.e., that none of its eigenvalues are zero such that $\tilde{\boldsymbol{A}}$ does not contain zeros on its diagonal.  Zero eigenvalues of $\boldsymbol{A}$ correspond to modes the do not propagate (they are neither leftgoing nor rightgoing).  Physically, these modes arise when the streamwise velocity is either sonic or zero and are therefore intrinsic to supersonic and wall-bounded flows \cite{Towne2015one}.  Mathematically, the zero eigenvalues make~(\ref{laplace transform of characteristic hyperbolic equation}) a differential-algebraic equation (DAE), and the operator $\boldsymbol{M}$ can no longer be defined as in~(\ref{definition of M}).  In the OWNS-P method, $\boldsymbol{M}$ is redefined by eliminating the states and equations associated with the zero eigenvalue modes \cite{Towne2021fast}.  This additional complication can be avoided in the OWNS-R formulation, as described below.
 
 Since the projection operator is applied to the state after each step in the march, the process of integrating the equations and forming the approximate projection operator can be considered separately in the OWNS-R method (unlike OWNS-P, where the projection is applied to the equation prior to integration, thus requiring that both steps be carefully considered together for singular $\boldsymbol{A}$).  First, (\ref{Equation: spatial marching equation}) is replaced with 
  \begin{equation} \label{Equation: spatial marching equation with singular A}
     \tilde{\boldsymbol{A}}\frac{d\hat{\boldsymbol{\phi}}}{dx}=\boldsymbol{L}\hat{\boldsymbol{\phi}}+ \hat{\boldsymbol{f}}_{\phi},
 \end{equation}
 to avoid taking the inverse of the singular matrix, where 
 \begin{equation}
   \boldsymbol{L}=-(s\boldsymbol{I}+\tilde{\boldsymbol{B}}).   
 \end{equation}
 Equation~(\ref{Equation: spatial marching equation with singular A}) is a DAE since $\tilde{\boldsymbol{A}}$ is singular.  The DAE can be integrated implicitly \cite{Hairer1996solving}; for example, using the implicit Euler method yields
  \begin{equation} \label{Equation: implicit Euler with singular A}
     (\tilde{\boldsymbol{A}}- \Delta x \boldsymbol{L})\hat{\boldsymbol{\phi}}(x_{n+1}) = \tilde{\boldsymbol{A}}\hat{\boldsymbol{\phi}}(x_n)+ \Delta x \hat{\boldsymbol{f}}_{\phi}.
 \end{equation}
 Since $\tilde{\boldsymbol{A}}-\Delta x \boldsymbol{L}$ is generally not singular, the state can be advanced as
 \begin{equation} \label{Equation: step integration for singular A}
      \hat{\boldsymbol{\phi}}(x_{n+1}) =(\tilde{\boldsymbol{A}}-\Delta x \boldsymbol{L})^{-1} (\tilde{\boldsymbol{A}}\hat{\boldsymbol{\phi}}(x_n)+ \Delta x \hat{\boldsymbol{f}}_{\phi}).
 \end{equation}

 Second, to form the approximate projection operator $\boldsymbol{P}_{N_{\beta}}$ without the need to invert the singular matrix $\tilde{\boldsymbol{A}}$, the definition of $\boldsymbol{Z}$ in (\ref{equation: formulation of approximate projection_Z}) is replaced with
\begin{equation} \label{Equation: Z term2}
        \boldsymbol{Z}=\prod\limits_{j=1}^{N_\beta}{[(\boldsymbol{L}-i{\beta_j}^{-}\tilde{\boldsymbol{A}})^{-1}(\boldsymbol{L}-i{\beta_j}^{+}\tilde{\boldsymbol{A}}) ]}.
\end{equation}
Then, the approximate projection operator $\boldsymbol{P}_{N_{\beta}}$ can be formed as before using (\ref{equation: formulation of approximate projection_P}).  Finally, applying this projection to the state vector obtained in (\ref{Equation: step integration for singular A}) completes the OWNS-R algorithm for singular ${\boldsymbol{A}}$.


\end{appendices}

 
 \bibliographystyle{model1-num-names}
 \bibliography{refs}

\end{document}